\documentclass[12pt]{article}
\renewcommand{\title}[1]{
\begin{center} \Large \bf #1 \end{center}
}
\renewcommand{\author}[2]{
 \begin{center} #1  \vspace{3mm} \\
  #2 \\
 \end{center}
\addvspace{\baselineskip}
}
\usepackage{amssymb}
\usepackage{amsmath}
\usepackage{multirow,bigdelim}
\usepackage{color}
\usepackage{ulem}

\usepackage{amsthm}
\newtheorem{thm}{Theorem}[section]
\newtheorem{prop}[thm]{Proposition}

\newtheorem{lem}[thm]{Lemma}
\newtheorem{fact}[thm]{Fact}
\newtheorem{ex}{Example}
\theoremstyle{definition}
\newtheorem{df}{Definition}

\theoremstyle{remark}
\newtheorem{rem}{Remark}

\theoremstyle{proof}
\newtheorem*{pf}{Proof}
\makeatletter
\@addtoreset{equation}{section}

\makeatother
\newcommand{\eq}[1]{(\ref{#1})} % equation number Eq.(#)
\usepackage{braket}
\date{\today}
\begin{document}

\baselineskip 5mm

\title{Hermitian-Einstein metrics \\
from noncommutative $U\left(1 \right)$ instantons
%
%Deformation Quantization of
%Gauge theories with Separation of Variables
%in Homogeneous K\"ahler Manifolds
}
%%%%%%%%%%%%%%%%%%%%%%%%%%%%%%%%%%%%%%%%%%%%%%%%%%%%%%%%%%%%%%%%

\author{${}^1$ Kentaro Hara,~ ${}^2$ Akifumi Sako and~ ${}^3$ Hyun Seok Yang}{
${}^1$ Department of Mathematics and Science Education\\
Tokyo University of Science,
1-3 Kagurazaka, Shinjuku-ku, Tokyo 162-8601, Japan\\
${}^2$   Department of Mathematics,\\
Tokyo University of Science,
1-3 Kagurazaka, Shinjuku-ku, Tokyo 162-8601, Japan\\
${}^3$ Center for Quantum Spacetime, Sogang University, Seoul 04107, Korea}

\abstract{
We show that Hermitian-Einstein metrics can be locally constructed by a map from (anti-)self-dual two-forms on Euclidean ${\mathbb R}^4$ 
to symmetric two-tensors
introduced in \cite{gi-u1-prl}. This correspondence is valid not only for a commutative space but also for a noncommutative space.
We choose $U(1)$ instantons on a noncommutative ${\mathbb C}^2$ as the self-dual two-form,
from which we derive a family of Hermitian-Einstein metrics.
We also discuss the condition when the metric becomes K\"ahler.}

\section{Introduction}

In this article, a linear map from differential two-forms to symmetric two-tensors
in two-dimensional Hermitian manifolds
introduced in \cite{gi-u1-prl} is studied.
The map reveals another aspect of Seiberg-Witten map.
The original Seiberg-Witten map is a map
from noncommutative gauge fields to commutative gauge fields with a background $B$-field \cite{sw-ncft}.
On the other hand, it has been interpreted in \cite{gi-u1-prl,gi-u1-plb,gi-u1-epl}
as a map from a noncommutative gauge field to a K\"ahler metric.\\

A purpose of this article is to clarify the map in \cite{gi-u1-prl,gi-u1-plb,gi-u1-epl}
which locally maps (anti-)self-dual two-forms on ${\mathbb C}^2$
to Hermitian-Einstein metrics of two-dimensional K\"ahler manifolds.
It might be worth noting that it is enough for these two-forms to be defined
as a symplectic structure on a commutative manifold,
although this map was developed in the context of Seiberg-Witten map in noncommutative gauge theory.
But this correspondence between the self-dual two-form and
Hermitian-Einstein metric can be lifted to noncommutative spaces
after (canonical or deformation) quantization \cite{ly-jkps2018}.\\

The second purpose of this article is to construct explicit examples of Hermitian-Einstein metrics from
noncommutative $U(1)$ instantons.
$U(1)$ instantons on noncommutative ${\mathbb C}^2$
were found by Nekrasov and Schwarz \cite{Nekrasov:1998ss}.
We will construct the two-form from a multi-instanton solution given in \cite{Ishikawa:2001ye}
where the noncommutative $U(1)$ instanton solutions are written in an operator form acting on a Fock space.
The Fock space is defined by Heisenberg algebra generated by noncommutative complex coordinates.
There is a dictionary between the linear operators acting on the Fock space
and usual functions \cite{Sako:2016gqb}. %[Sako-Umetsu]
The dictionary %[Sako-Umetsu]
is applicable for arbitrary noncommutative K\"ahler manifold
obtained by deformation quantization
with separation of variables \cite{Karabegov1996}.
Concrete Hermitian-Einstein metrics are obtained by translating the noncommutative instantons as linear operators into ordinary functions by using
the dictionary in \cite{Sako:2016gqb}.\\

The third purpose is to clarify the K\"ahler condition for the metrics derived
from noncommutative $U(1)$ instantons.
Since a K\"ahler manifold is a symplectic manifold too although the reverse
is not necessarily true, one can quantize the K\"ahler manifold by quantizing
a Poisson algebra derived from the underlying symplectic structure of the K\"ahler geometry,
as recently clarified in \cite{ly-jkps2018}.
We will show that the metric derived from noncommutative $U(1)$ instantons
becomes a K\"ahler metric if the underlying Poisson algebra of $U(1)$ instantons or its quantization is an associative algebra. \\

Here we mention some studies related with subjects of this article.
It has been conjectured in \cite{inov,ideal-sheaf} that
NC $U(1)$ gauge theory is the fundamental description
of K\"ahler gravity at all scales including the Planck scale
and provides a quantum gravity description such as quantum gravitational foams.
Recently it was shown in \cite{hsy-jhep09,review4,hsy-jpcs12} that
the electromagnetism in noncommutative spacetime can be realized
as a theory of gravity and the symplectization of spacetime geometry is the origin of gravity.
Such picture is called emergent gravity and it proposes a candidate of the origin of spacetime.
See also related works in Refs. \cite{rivelles,review1,review2,review3,yasi-prd10,lee-yang,review6,review7,kawai-ks2016}
As a bottom-up approach of the emergent gravity formulated in \cite{our-jhep12},
the Eguchi-Hanson metric \cite{eh-plb,eh-ap} in four-dimensional Euclidean gravity
is used to construct anti-self-dual symplectic $U(1)$ gauge fields,
and $U(1)$ gauge fields corresponding to the Nekrasov-Schwarz instanton \cite{Nekrasov:1998ss}
are reproduced by the reverse process \cite{Lee:2012rb}. %{sty-06,prl-06}.
As a top-down approach of emergent gravity, the $U(1)$ instanton
found by Braden and Nekrasov \cite{bn-inst} derives a corresponding gravitational metric.\\

The organization of this paper is as follows.
In section \ref{sect2},
some linear algebraic formulas for self-duality are prepared.
In section \ref{Einstein and ASD},
the correspondence between the self-dual two-forms and Hermitian-Einstein metrics is studied.
In section \ref{sect4},
Hermitian-Einstein metrics are explicitly constructed from noncommutative $U(1)$ instantons.
In section \ref{sect5}, the gauge theory realization of the K\"ahler condition is studied.
In section \ref{sect6},
we discuss an outlook of this subject.
Some technical details are left for the appendices.

\section{Self-duality}\label{sect2}

\begin{df}[Hodge star operator]
An automorphism $\star$ on the set of $4\times 4$ alternative matrices is defined as
\[\star\left[\left(
\begin{array}{cccc}
0 &\omega_{12} &\omega_{13} & \omega_{14} \\
-\omega_{12} &0 &\omega_{23} & \omega_{24} \\
-\omega_{13} &-\omega_{23} &0 & \omega_{34} \\
-\omega_{14} & -\omega_{24}&-\omega_{34} & 0
\end{array}
\right) \right]:=\left(
\begin{array}{cccc}
0 &\omega_{34} &-\omega_{24} & \omega_{23} \\
-\omega_{34} &0 &\omega_{14} & -\omega_{13} \\
\omega_{24} &-\omega_{14} &0 & \omega_{12} \\
-\omega_{23} & \omega_{13}&-\omega_{12} & 0
\end{array}
\right),\]
\[\left(i.e., ~ \omega_{12}\leftrightarrow \omega_{34}, ~ \omega_{13}\leftrightarrow \omega_{42}, ~ \omega_{14}\leftrightarrow \omega_{23}\right). \]
%\sout{and $\star\omega_{kl}$ is defined as} \textcolor[rgb]{0.00,1.00,0.00}{(Th%e below seems to be superfluous)
%\[\left(
%\begin{array}{cccc}
%0 &\star\omega_{12} &\star\omega_{13} &\star \omega_{14} \\
%-\star\omega_{12} &0 &\star\omega_{23} & \star\omega_{24} \\
%-\star\omega_{13} &-\star\omega_{23} &0 & \star\omega_{34} \\
%-\star\omega_{14} & -\star\omega_{24}&-\star\omega_{34} & 0
%\end{array}
%\right):=\star\left[\left(
%\begin{array}{cccc}
%0 &\omega_{12} &\omega_{13} & \omega_{14} \\
%-\omega_{12} &0 &\omega_{23} & \omega_{24} \\
%-\omega_{13} &-\omega_{23} &0 & \omega_{34} \\
%-\omega_{14} & -\omega_{24}&-\omega_{34} & 0
%\end{array}
%\right) \right].\]}
In other words, $\star\omega_{kl}$ is defined as
\[\star\omega_{kl}
%\textcolor[rgb]{0.00,1.00,0.00}{(=\sum _{m,n}^4\frac{\varepsilon_{klmn}\omega_{%mn}}{2})}
=\frac{1}{2} \sum _{m,n}^4 \varepsilon_{klmn} \omega_{mn},\] where $\varepsilon_{klmn}$ is Levi-Civita symbol.
The operator $\star$ is called the Hodge star operation in Euclidean $\mathbb{R}^4$.
\end{df}

\begin{df}[Anti-self-dual matrix]
A $4\times 4$ alternative matrix $\omega^{\pm}$ is an (anti-)self-dual matrix if
\begin{align}
\star\omega^{\pm}=\pm \omega^{\pm}. \label{asd}
\end{align}
\end{df}

An (anti-)self-dual matrix $\theta^{\pm}$ is defined as
\begin{align}
\theta^{\pm }:=\left(
\begin{array}{cccc}
0 &-\theta & 0& 0 \\
\theta &0 &0 & 0 \\
0 &0 &0 & \mp \theta \\
0 &0 & \pm \theta & 0
\end{array}
\right) & \label{theta}
\end{align}

where $\theta$ is a real number.
Note that $\omega^{\pm}$ and $\theta^{\mp}$ commute each other:
\begin{align}
\omega^{\pm}\theta^{\mp}=\theta^{\mp}\omega^{\pm}. \label{omegatheta}
\end{align}

\begin{df}[Matrix $g^{\pm}$]Let $E_4$ be the $4\times4$ unit matrix and $\omega^{\pm}$
be a $4\times 4$ (anti-)self-dual matrix.
Assume that $\det\left[E_4-\omega^{\pm}\theta^{\mp} \right]\neq 0$,
then $4\times 4$ matrix $g^{\pm}$ is defined as
\[g^{\pm}:=2\left(E_4-\omega^{\pm}\theta^{\mp} \right)^{-1}-E_4. \label{g-metric}
%,G^{\pm}:=\left(E_4-\omega\theta^{\mp} \right)^{-1}
%R_{\bar{j}k}:=\partial _{\bar{j}}\partial _{k}\log \left|g\right|
\]
\end{df}

\begin{rem}\label{gsym}$g^{\pm}$ is a symmetric matrix because of (\ref{omegatheta})
and it can be inverted to
\[\omega^{\pm}
=\left(g^{\pm} - E_4 \right) \left(g^{\pm} + E_4 \right)^{-1}
\left(\theta^\mp \right)^{-1}.\]
\end{rem}
The Remark \ref{gsym} allows us to regard $g^{\pm}$
as a metric tensor since it is symmetric and nondegenerate.

\begin{lem}\label{lem2.1} For any $4\times 4$ (anti-)self-dual matrix $\omega^\pm$,
\begin{align}\label{det=1}
\star\omega^\pm =\pm \omega^\pm \Longrightarrow \det\left[ g^\pm \right]=1.
\end{align}
\end{lem}
This lemma is proved by a direct calculation.

\begin{df}\label{iskew}The map $\iota_{skew}:\left\{\omega_{\mathbb{C}}\in M_2[\mathbb{C}]
\: |~ \omega_{\mathbb{C}}^\dagger=-\omega_{\mathbb{C}} \right\}\longrightarrow M_4[\mathbb{R}]$
is defined as \[\iota_{skew}\left[\left(
\begin{array}{cc}
\omega_{\mathbb{C}1\bar{1}} & \omega_{\mathbb{C}1\bar{2}} \\
\omega_{\mathbb{C}2\bar{1}} & \omega_{\mathbb{C}2\bar{2}}
\end{array}
\right) \right]=\left(
\begin{array}{cccc}
0 &2\mathrm{i}\omega_{\mathbb{C}1\bar{1}} & \omega_{\mathbb{C}1\bar{2}}-\omega_{\mathbb{C}2\bar{1}}
& \mathrm{i}\left(\omega_{\mathbb{C}1\bar{2}}+\omega_{\mathbb{C}2\bar{1}} \right) \\
-2\mathrm{i}\omega_{\mathbb{C}1\bar{1}}  &0 &-\mathrm{i}\left(\omega_{\mathbb{C}1\bar{2}}+\omega_{\mathbb{C}2\bar{1}} \right)
& \omega_{\mathbb{C}1\bar{2}}-\omega_{\mathbb{C}2\bar{1}}  \\
-\omega_{\mathbb{C}1\bar{2}}+\omega_{\mathbb{C}2\bar{1}} &\mathrm{i}\left(\omega_{\mathbb{C}1\bar{2}}+\omega_{\mathbb{C}2\bar{1}} \right) &0
& 2\mathrm{i}\omega_{\mathbb{C}2\bar{2}}  \\
-\mathrm{i}\left(\omega_{\mathbb{C}1\bar{2}}+\omega_{\mathbb{C}2\bar{1}} \right)
&-\omega_{\mathbb{C}1\bar{2}}+\omega_{\mathbb{C}2\bar{1}} &-2\mathrm{i}\omega_{\mathbb{C}2\bar{2}}  & 0
\end{array}
\right).\]
Note that $\omega_{\mathbb{C}1\bar{1}}$ and $\omega_{\mathbb{C}2\bar{2}}$ are pure imaginary.
\end{df}
If the coordinate transformation on the coordinate neighborhood is $z_1:=x^2+\mathrm{i}x^1,z_2:=x^4+\mathrm{i}x^3$, then
the $\iota_{skew}$ is the pull-back of a two-form. This means

\[ \sum _{k,l=1}^2\omega_{\mathbb{C}k\bar{l}}\mathrm{d}z_k \wedge \mathrm{d}\bar{z}_l
=\frac{1}{2}\sum _{k,l=1}^4\omega_{kl}\mathrm{d}x^k \wedge \mathrm{d}x^l=
\frac{1}{2}\sum _{k,l=1}^4\left(\iota_{skew}\left[ \omega_\mathbb{C}\right] \right)_{kl}\mathrm{d}x^k \wedge \mathrm{d}x^l.\]
The above $\iota_{skew}$ is defined as satisfying this relation.

\begin{rem}$\iota_{skew}$ satisfies the following relation
\[\det\left[\iota_{skew}\left[\omega_{\mathbb{C}}\right] \right]=16\left(\det\left[\omega_{\mathbb{C}} \right]\right)^2.\]
\end{rem}

Using this result, the following lemma can be deduced.
\begin{lem}\label{lem2.2}
Suppose that the anti-Hermitian matrix $\omega_{\mathbb{C}}$ satisfies $\omega_{\mathbb{C}2\bar{2}}=-\omega_{\mathbb{C}1\bar{1}}$,
i.e. ${\rm tr \omega_{\mathbb{C}}} = 0$.
Then the two-form $\iota_{skew} [\omega_{\mathbb{C}}]$ is anti-self-dual, i.e.,
\[\star\left\{ \iota_{skew}\left[\left(
\begin{array}{cc}
\omega_{\mathbb{C}1\bar{1}} & \omega_{\mathbb{C}1\bar{2}} \\
\omega_{\mathbb{C}2\bar{1}} & \omega_{\mathbb{C}2\bar{2}}
\end{array}
\right)\right]\right\}=-\iota_{skew}\left[\left(
\begin{array}{cc}
\omega_{\mathbb{C}1\bar{1}} & \omega_{\mathbb{C}1\bar{2}} \\
\omega_{\mathbb{C}2\bar{1}} & \omega_{\mathbb{C}2\bar{2}}
\end{array}
\right)\right].\]
\end{lem}

\section{Hermitian-Einstein metrics and (anti-)self-dual two-forms}\label{Einstein and ASD}
In this section, we discuss how to make a Hermitian-Einstein metric from an anti-self-dual two-form.
Let us define a $u\left(1 \right)$-valued two-form on $\mathbb{R}^4$ by
\[
%\textcolor[rgb]{0.98,0.00,0.00}{\omega = \frac{1}{2}}
 \sum _{k,l=1}^4\omega_{kl}\mathrm{d}x^k \wedge \mathrm{d}x^l.\]
where $\omega$ is an alternative matrix $\left(\omega \right)_{kl}:=\omega_{kl}$.
If $\omega$ is an anti-self-dual matrix, then the two-form is called anti-self-dual two-form.

%Instanton is an example of the anti-selfdual two-form.
%Let $F^-$ be the curvature form of the principal bundle $P\rightarrow M.$ If $F^-$ is the anti-selfdual two-form
%it is called instanton. The instanton is a solution of the Yang-Mills equations on $\mathbb{R}^4$.
%\[\star \mathrm{d}\star F^-=0.\]
%\textcolor[rgb]{0.00,1.00,0.00}{\sout{In the following of this article 4-dimens%ional Riemannian
%manifolds are studied.}}

\subsection{Ricci flat metrics and Hermitian-Einstein metrics}

Let $M$ be a Hermitian manifold and $h$ be its metric.
%The Riemannian curvature of a Hermitian manifold $M$ is defined as
%\[{R_{i\bar{j}k}}^l =\partial_i\Gamma^l_{\bar{j}k}  - \partial_{\bar{j}}\Gamma^%l_{ik}
%+\Gamma^n_{\bar{j}k} \Gamma^l_{in} -\Gamma^n_{ik}\Gamma^l_{\bar{j}n}.\]
%Here $\Gamma^l_{\bar{j}k}$ is the Christoffel symbol.
%For a Hermitian manifold, the Christoffel symbols are given by
%\[\Gamma^l_{jk}=h^{l\bar{q}}\frac{\partial h_{j\bar{q}} }{\partial z^k}.\]
%The curvature tensor of a Hermitian manifold $M$ is obtained as
%\[
%R_{i\bar{j} k \bar{l}}= -\frac{\partial^2 h_{\bar{j}i}}{\partial z^k\partial \b%ar{z}^l}
% +h^{p\bar{q}}\frac{\partial h_{i\bar{q}} }{\partial z^k}\frac{\partial h_{\bar%{j}p}}{\partial \bar{z}^l}.
%\]
As a well-known fact, Ricci curvature $R_{\bar{j}k}$ for a Hermitian manifold $\left(M,h,\nabla  \right)$
with the Levi-Civita connection $\nabla$ takes a simple form
\begin{align}\label{ricci}
R_{\bar{j}k}=\partial _{\bar{j}}\partial _{k}\log \left(\det\left[ h\right] \right).
\end{align}
%Assume $\left(M,h,\nabla  \right)$ be an Hermitian manifold with Levi-Civita connection,
%$R_{\bar{k}l}$ be Ricci curvatures, $h_{\bar{k}l}$ be metric tensors and
See, for example, \cite{Kobayashi_Nomizu,besse}.
Let $\lambda$ be a cosmological constant. When $h$ satisfies the Einstein's equation.
\[R_{\bar{k}l}=\lambda h_{\bar{k}l}\]
then $M$ is called an Einstein manifold. In this paper we will focus on a Ricci flat manifold (i.e.
$R_{\bar{k}l} = 0$ or $\lambda = 0$).
%Let $R$ be a scalar curvature. When $h$ satisfies the Einstein's equation \[R_{\bar{k}l}-\frac{1}{2}Rh_{\bar{k}l}=0,\]
%then $M$ is called an Einstein manifold.If $M$ is a Ricci flat manifold (i.e. $R_{\bar{k}l}=0 $), then $M$ is an Einstein manifold.
We consider $M$ as a real manifold with local coordinates $x^\mu\left(\mu=1,2,3,4 \right)$.
%Let $M$ be a 2-dim Hermitian manifold.$U\xrightarrow{\varphi} \mathbb{C}^2\xrightarrow{\sim }\mathbb{R}^4$.
\begin{df}\label{isym}The map $\iota_{sym}:\left\{h\in M_2[\mathbb{C}]~|~h^\dagger=h \right\}\longrightarrow M_4[\mathbb{R}]$ is defined as
\[\iota_{sym}\left[\left(
\begin{array}{cc}
h_{1\bar{1}} & h_{1\bar{2}} \\
h_{2\bar{1}} & h_{2\bar{2}}
\end{array}
\right) \right]
=\left(
\begin{array}{cccc}
h_{1\bar{1}} &0 &\frac{1}{2}\left(h_{1\bar{2}}+h_{2\bar{1}} \right) &\frac{1}{2\mathrm{i}}\left(h_{2\bar{1}}-h_{1\bar{2}} \right)  \\
0 &h_{1\bar{1}} &-\frac{1}{2\mathrm{i}}\left(h_{2\bar{1}}-h_{1\bar{2}} \right) & \frac{1}{2}\left(h_{1\bar{2}}+h_{2\bar{1}} \right)  \\
\frac{1}{2}\left(h_{1\bar{2}}+h_{2\bar{1}} \right)  &-\frac{1}{2\mathrm{i}}\left(h_{2\bar{1}}-h_{1\bar{2}} \right) &h_{2\bar{2}} &0  \\
\frac{1}{2\mathrm{i}}\left(h_{2\bar{1}}-h_{1\bar{2}} \right) &\frac{1}{2}\left(h_{1\bar{2}}+h_{2\bar{1}} \right)  &0 & h_{2\bar{2}}
\end{array}
\right).\]
where $h$ is a matrix and $\left( h\right)_{k\bar{l}}:=h_{k\bar{l}}$.
\end{df}

\begin{rem}Assume that $h$ is a Hermitian metric.
If the coordinate transformation on a coordinate neighborhood is $z^1:=x^2+\mathrm{i}x^1,z^2:=x^4+\mathrm{i}x^3$,
the $\iota_{sym}$ is then the pull-back of the Hermitian metric given by
\[ \sum _{k,l=1}^2h_{k\bar{l}}\mathrm{d}z_k \mathrm{d}\bar{z}_l=
\sum _{k,l=1}^4\left(\iota_{sym}\left[ h\right] \right)_{kl}\mathrm{d}x^k \mathrm{d}x^l.\]
Hence $\iota_{sym}$ squares the determinant:
\[\det\left[\iota_{sym}\left(h \right) \right]=(\det \left[h \right])^2.\]
\end{rem}

A Hermitian metric made with $\iota_{sym}^{-1}$ will be used below.\\
%\begin{df}$\star :\mathfrak{g}\otimes \Lambda^2\left(M \right)\longrightarrow \mathfrak{g}\otimes \Lambda^2\left(M \right)$ is defined as
%\[\star \left( \sum _{k,l=1}^{dim M}\omega_{k\bar{l}}\mathrm{d}z^k \wedge \mathrm{d}\bar{z}^l\right)
%:=\sum _{k,l=1}^{dim M}\star \omega_{k\bar{l}}\mathrm{d}z^k \wedge \mathrm{d}\bar{z}^l \]\end{df}

\begin{df}If $\tilde{h}\in C^\infty \left(U,M_2[\mathbb{C}] \right)$ and $\tilde{h}^\dagger =\tilde{h}$, then
$$\tilde{h}>0~in~U ~~ \Longleftrightarrow ~~ \forall u\in U, ~ \tilde{h}\left(u \right)>0 $$
where $\tilde{h}\left(u \right)>0$ means that $\tilde{h}$ is positive definite as a Hermitian matrix.
\end{df}

\begin{lem} \label{lem3.1}
If $h \in C^\infty \left(U,M_2[\mathbb{C}] \right)$ is a Hermitian matrix with $\det\left[h \right]=1$ and
$h$ is positive (negative) at ${}^\exists p\in U$, then $h$ is positive (negative) in $U$.
\end{lem}
\begin{pf}This follows from
\begin{align*}
\lefteqn{\left\{h\in M_2[\mathbb{C}]~\big|~h=h^\dagger,~\det\left[h \right]=1 \right\}} \\
 &=\left\{\left(
\begin{array}{cc}
a & b \\
\bar{b} & d
\end{array}
\right)
\in M_2[\mathbb{C}]~\big|~a,d\in\mathbb{R},~a>0,~d>0,~a d\geq 1,~\left|b \right|=\sqrt{a d- 1} \right\}\\
&\coprod \left\{\left(
\begin{array}{cc}
a & b \\
\bar{b} & d
\end{array}
\right)
\in M_2[\mathbb{C}]~\big|~a,d\in\mathbb{R},~a<0,~d<0,~a d\geq 1,~\left|b \right|=\sqrt{a d- 1} \right\}
\end{align*}
which means two spaces are disconnected.
\qed\end{pf}

From the above discussions, the following theorem is obtained.
\begin{thm} \label{thm-kal-ma}
Let $\omega^\pm$ be an (anti-)self-dual two-form on an open neighborhood $U$, i.e.
$\star \omega^\pm =\pm\omega^\pm$,
\label{masspro}
and
\begin{align} \label{sdmetric}
h^\pm:=\iota_{sym}^{-1} \left[2 \left(E_4-\omega^\pm \theta^\mp  \right)^{-1}-E_4 \right].
\end{align}
Then $h^\pm$ gives a Ricci-flat Hermitian metric on $U$.
%\[R_{\bar{j}k}=\partial _{\bar{j}}\partial _{k}\log \left|h\right|=0\]
So $(U, h^\pm)$ is a local realization of an Einstein manifold.
\end{thm}

\begin{pf}Because of Lemma \ref{lem2.1},
if $\star \omega^\pm =\pm\omega^\pm$, then
\begin{align}\label{sd}
\det\left[h^\pm  \right]=1.
\end{align}
%~\left(h^\pm \right)^\dagger=h^\pm.\]
Because of Lemma \ref{lem3.1} and Remark \ref{gsym}, $h^\pm$ is a metric tensor.
From equations (\ref{ricci}) and (\ref{sd}),
$R_{\bar{j}k}=\partial _{\bar{j}}\partial _{k}\log\left(\det\left[ h^\pm\right] \right)=0$.
\qed\end{pf}

%%%%%%%%%%%%%%
%\begin{pf}By the Lemma \ref{lem2.1} and Remark \ref{gsym}, if $\star \omega^\pm% =\pm\omega^\pm$, then
%\[\det\left[h^\pm  \right]=1,~\left(h^\pm \right)^\dagger=h^\pm.\]
%From (\ref{ricci}), $R_{\bar{j}k}=\partial _{\bar{j}}\partial _{k}\log\left(\de%t\left[ h^\pm\right] \right)=0$.
%\qed\end{pf}
%%%%%%%%%%%%%%%%%
%From Lemma \ref{sym}, $h$ is a Hermitian metric.

Local complex coordinates can be arranged in such a way that the Jacobians of the
transition functions on overlapping charts are one on all the overlaps.
In that case, $\det [h^\pm]$ is a globally defined function and
the Ricci-flat condition reduces to the Monge-Amp\'ere equation \cite{ma-eq}
\begin{equation}\label{ma-eq}
\det [h^\pm] = \kappa,
\end{equation}
where the constant $\kappa$ is related to the volume of a K\"ahler manifold that depends only on the K\"ahler
class. Therefore Theorem \ref{thm-kal-ma} implies that the self-duality for the two-form $\omega^\pm$ is equivalent to the Ricci-flat condition \eq{ma-eq} of K\"ahler manifolds defined
by the metric $h^\pm$ \cite{u1-cym}.

%%%%%%%%%%%%%%%%%%%%%%%%%%%%%%%%%%%%%%%%%%%%%%%%%%%%%%%%%%%%%%%%%%%%%%%%%%%%%%%%%%%%%%%%%%%%%%%%%%%%%
\section{Hermitian-Einstein metric from noncommutative instanton on $\mathbb{C}^2$}\label{sect4}
In the previous section we found the way to construct a Hermitian-Einstein metric from an (anti-)self-dual two-form.
To construct the Hermitian-Einstein metric, we will employ the instanton curvature
on noncommutative $\mathbb{C}^2$ as the (anti-)self-dual two-form.
There are many ways to obtain noncommutative $\mathbb{C}^2$
(see \cite{nc-review} for a review and references therein).
We use the Fock representation of noncommutative $\mathbb{C}^2$ given in \cite{Sako:2016gqb},
which is based on the Karabegov's deformation quantization \cite{Karabegov1996}.
There is a simple dictionary between the Fock representation and ordinary functions.
Using the dictionary, the Hermitian-Einstein metric is expressed in terms of usual functions.

\subsection{Noncommutative $\mathbb{C}^2$}

Consider a noncommutative algebra $\left(C^\infty \left(\mathbb{C}^2 \right)\left[\left[\hbar \right] \right],* \right)$ led by (\ref{f*g})
in Appendix \ref{dq}.
The star product induces a Heisenberg algebra
\begin{align}
\label{ncp}
\left[z^k,\bar{z}^l\right]_*=-\zeta_k\delta_{kl},
\qquad \left[z^k,z^l\right]_*=0,\qquad \left[\bar{z}^k,\bar{z}^l\right]_*=0 ,
\end{align}
where $\left[x,y\right]_*:=x*y-y*x $. We represent it by creation and annihilation operators given by
\[a_k:=\frac{\bar{z}^k}{\sqrt{\zeta_k}},\qquad a_k^\dagger:=\frac{z^k}{\sqrt{\zeta_k}},\]
 then
\[\left[a_k,a_l^\dagger \right]_*=\delta_{kl},\qquad \left[a_k^\dagger ,a_l^\dagger \right]_*=0,\qquad \left[a_k,a_l\right]_*=0.\]
%There is an embedding from the algebra to the linear space$\left\{C^\infty \left(\mathbb{C}^2 \right) \right\}\left(\left(\hbar \right) \right)$.
In the following $\zeta_1=\zeta_2=\zeta>0$ is assumed.\\

Note that the choice of a noncommutative parameter has the freedom associated
with a choice of a background two-form \cite{sw-ncft}.
Here the $\zeta$ in (\ref{ncp}) is regarded as the only noncommutative parameter.
However, in Section \ref{sect5}, we will implicitly assume the identification $\zeta := 2 \theta$
since we will work in the background-independent prescription, i.e. $\theta = B^{-1}$.\\

The algebra $\mathcal{F}$ on $\mathbb{C}$ is defined as follows.
The Fock space  $\mathcal{H}$ is  a linear space
spanned by the bases generated by acting $a_l^\dagger$'s on $\Ket{0,0}$ :
\begin{align}
\frac{1}{\sqrt{m_1!m_2!}}\left( a_1^\dagger\right)^{m_1}_**\left( a_2^\dagger\right)^{m_2}_*\Ket{0,0}= \Ket{m_1,m_2} ,\label{fsp}
\end{align}
where $m_1$ and $m_2$ are positive integers and $\left( a\right)^m_*$ stands for $\overbrace{a * \cdots  * a}^m$.
The ground state $\Ket{0,0}$ satisfies $a_l\Ket{0,0}=0, ~ \forall  ~ l$.
Here, we define the basis of a dual vector space by acting $a_l$'s on $\Bra{0,0}$ as
$$\frac{1}{\sqrt{n_1!n_2!}}\Bra{0,0}\left(a_1\right)^{n_1}_**\left(a_2\right)^{n_2}_*=\Bra{n_1,n_2} , $$
where $\Bra{0,0}$ satisfies $\Bra{0,0}a_l^\dagger=0, ~ \forall  ~ l$. Then we define a set of linear operators as
\begin{align}
\mathcal{F}:=span_{\mathbb{C}}\left(\Ket{m_1,m_2}\Bra{n_1,n_2};m_1,m_2,n_1,n_2=0,1,2,\cdots \right) \label{ketbra}
\end{align}
where $\left( \Ket{m_1,m_2}\Bra{n_1,n_2}\right)\Ket{k_1,k_2}=\delta_{k_1n_1}\delta_{k_2n_2}\Ket{m_1,m_2}$ and
$\Bra{k_1,k_2}\left( \Ket{m_1,m_2}\Bra{n_1,n_2}\right)=\delta_{k_1m_1}\delta_{k_2m_2}\Bra{n_1,n_2}$.
The product on $\mathcal{F}$ is defined as
$$\left(\Ket{j_1,j_2}\Bra{k_1,k_2} \right)\circ\left( \Ket{m_1,m_2}\Bra{n_1,n_2}\right):=
\delta_{k_1m_1}\delta_{k_2m_2}\Ket{j_1,j_2}\Bra{n_1,n_2},$$
so, $\mathcal{F}$ is an algebra.

There is a one to one correspondence between $\mathcal{F}$ and some subalgebra of $C^\infty \left(\mathbb{C}^2 \right)$.
For arbitrary noncommutative K\"ahler manifold
obtained by deformation quantization
with separation of variables \cite{Karabegov1996},
we can find the similar correspondence \cite{Sako:2016gqb}.
The following is the simplest example of the correspondence.
\begin{df}\label{iota}(Twisted Fock representation). The linear map
$\iota:\mathcal{F}\longrightarrow C^\infty \left(\mathbb{C}^2 \right)$
is defined as
\begin{align}
\iota\left(\Ket{m_1,m_2}\Bra{n_1,n_2} \right)=e_{\left( m_1,m_2,n_1,n_2\right)}:
=\frac{z_1^{m_1}z_2^{m_2}\mathrm{e}^{-\frac{z^1\bar{z}^1+z^2\bar{z}^2}{\zeta}}\bar{z}_1^{n_1}
\bar{z}_2^{n_2}}{\sqrt{m_1!m_2!n_1!n_2!}\left( \sqrt{\zeta}\right)^{m_1+m_2+n_1+n_2}},\label{Fockrep}
\end{align}
especially $\iota\left( \Ket{0,0}\Bra{0,0}\right)=e_{\left( 0,0,0,0\right)}=\mathrm{e}^{-\frac{z^1\bar{z}^1+z^2\bar{z}^2}{\zeta}}.$
\end{df}

\begin{prop}Let $\iota \left(\mathcal{F}\right)$ be defined by
\begin{align}
\iota \left(\mathcal{F}\right):=span_{\mathbb{C}}\left(e_{\left( m_1,m_2,n_1,n_2\right)};m_1,m_2,n_1,n_2=0,1,2,\cdots \right).
\end{align}
Then $\left\{\iota \left(\mathcal{F}\right),* \right\}$ is an algebra where $*$ is in (\ref{f*g}).
\end{prop}
\begin{pf}
After a little algebra, one can deduce the following identity
\begin{align}
e_{\left( k_1,k_2,l_1,l_2\right)}*e_{\left( m_1,m_2,n_1,n_2\right)}=\delta_{l_1m_1}\delta_{l_2m_2}e_{\left( k_1,k_2,n_1,n_2\right)}. \label{e*e}
\end{align}
Details are given in \cite{Sako:2016gqb}.
\qed\end{pf}
The identity (\ref{e*e}) derives the following fact.
\begin{prop}\label{dictionary}The algebras $\left(\mathcal{F},\circ\right)$ and $\left\{ \iota \left(\mathcal{F}\right),*\right\}$
are isomorphic.
\end{prop}

This isomorphism $\iota$ is a ``Fock space - function space'' dictionary.
From this isomorphism, we do not distinguish these two algebras and we only use $*$ to represent products in the following.\\
\bigskip

%\begin{fact}\[\iota\left[\left( \Ket{k_1,k_2}\Bra{l_1,l_2}\right)\left( \Ket{m_1,m_2}\Bra{n_1,n_2}\right) \right]
%=\iota\left[\left( \Ket{k_1,k_2}\Bra{l_1,l_2}\right) \right]* \iota\left[\left( \Ket{m_1,m_2}\Bra{n_1,n_2}\right) \right]\]\end{fact}

%\subsection{Noncommutative gauge theory}
Here we consider a $U(1)$ gauge theory on noncommutative $\mathbb{C}^2$.
$U(1)$ gauge connection in the noncommutative space is defined as follows (see for example \cite{Nekrasov:2000ih}).

\begin{df}Rescaled coordinates of $\mathbb{C}^2$ are defined as
\[\hat{\partial }_{z_l}:=\frac{\bar{z}_l}{\zeta_l}.\]
This acts on $\mathcal{H}$ as a linear operator.
\end{df}

Using $\hat{\partial }_{z_l},\hat{\partial }_{\bar{z}_m}$,
let us introduce covariant derivatives and the gauge curvature as follows.

\begin{df}\label{cov}Covariant derivatives for a scalar field in fundamental representation $\phi\in \mathcal{F}$ on noncommutative $\mathbb{C}^2$ are defined as
\[\hat{\nabla }_{z_l}\hat{\phi}:=\left[\hat{\partial }_{z_l},\hat{\phi }\right]_*+\hat{A}_{z_l}*\hat{\phi}
=-\hat{\phi}*\hat{\partial }_{z_l}+\hat{D}_{z_l}*\hat{\phi}\]
where we define a local gauge field $\hat{A}_{z_l}\in \mathcal{F}$ and
\[\hat{D}_{z_l}:=\hat{\partial }_{z_l}+\hat{A}_{z_l}.\]
The gauge curvature is defined as
\begin{align}
\hat{F}_{z_l\bar{z}_m}:&=\mathrm{i}\left[\hat{\nabla }_{z_l},\hat{\nabla }_{\bar{z}_m} \right]_*
=-\frac{\mathrm{i}\delta_{lm}}{\zeta_l}+\mathrm{i}\left[\hat{D}_{z_l},\hat{D}_{\bar{z}_m} \right]_*, \\
\hat{F}_{z_lz_m}:&=\mathrm{i}\left[\hat{\nabla }_{z_l},\hat{\nabla }_{z_m} \right]_*
=\mathrm{i}\left[\hat{D}_{z_l},\hat{D}_{z_m} \right]_*, \nonumber  \\
\hat{F}_{\bar{z}_l\bar{z}_m}:&=\mathrm{i}\left[\hat{\nabla }_{\bar{z}_l},\hat{\nabla }_{\bar{z}_m} \right]_*
=\mathrm{i}\left[\hat{D}_{\bar{z}_l},\hat{D}_{\bar{z}_m} \right]_*. \nonumber
\end{align}
\end{df}

\subsection{Ricci-flat metrics from noncommutative $k$-instantons}
In this section, we make Ricci-flat metrics on a local neighborhood from noncommutative instantons on ${\mathbb C}^2$.
As we saw in Section \ref{Einstein and ASD}, (anti)-self-dual two-forms satisfying (\ref{asd}) derive Ricci-flat metrics.
Nekrasov and Schwarz found in \cite{Nekrasov:1998ss} how to construct noncommutative instantons on ${\mathbb C}^2$
by using the ADHM method and the general solutions for the $U(1)$ gauge theory are given in \cite{Nekrasov:2000ih}.
We introduce the commutation relation of complex coordinates as (\ref{ncp}).
As (anti)-self-dual two-forms in Section \ref{Einstein and ASD},
we employ noncommutative instantons given in \cite{Ishikawa:2001ye}.\\

The general instanton solutions (see \cite{Ishikawa:2001ye}) satisfy the (anti)-self-dual relation.
An instanton curvature tensor is described by
\[\hat{F}^-_{\mathbb{C}}\left[k \right]:=\left(
\begin{array}{cc}
\hat{F}^-_{z_1\bar{z}_1}\left[k \right] & \hat{F}^-_{z_1\bar{z}_2}\left[k \right] \\
\hat{F}^-_{z_2\bar{z}_1}\left[k \right] & -\hat{F}^-_{z_1\bar{z}_1}\left[k \right]
\end{array}
\right) , \]
and satisfies (\ref{asd}):
\begin{align}
\star\left(\iota_{skew}\left(\hat{F}^-_{\mathbb{C}}\left[k \right] \right) \right)=-\iota_{skew}\left(\hat{F}^-_{\mathbb{C}}\left[k \right] \right).
\end{align}
See Lemma \ref{lem2.2} in Section \ref{sect2}.
%Here $\hat{F}^-_{\mathbb{C}}\left[k \right]$ is an instanton curvature and $\iota_{skew} \left(\hat{F}^-_{\mathbb{C}}\left[k \right] \right)$ satisfies .
This fact leads to the following result.
\begin{prop}
If $\hat{F}^-_{\mathbb{C}}$ is a $k$-instanton curvature tensor of $U(1)$ gauge theory
on noncommutative ${\mathbb C}^2$,
and
\begin{align}
h\left[k \right]:= & \iota^{-1}_{sym}\left\{
2\left(E_4-\iota_{skew}\left(\hat{F}^-_{\mathbb{C}}\left[k \right] \right) \theta^+ \right)^{-1}-E_4 \right\} \nonumber \\
=&\frac{1}{4\left|\hat{F}_\mathbb{C}^-\left[k \right] \right|\theta^2-1}\left(
\begin{array}{cc}
-4\mathrm{i}\hat{F}^-_{z_1\bar{z}_1}\left[k \right]\theta-2 & -4\mathrm{i}\hat{F}^-_{z_1\bar{z}_2}\left[k \right]\theta \\
-4\mathrm{i}\hat{F}^-_{z_2\bar{z}_1}\left[k \right]\theta & 4\mathrm{i}\hat{F}^-_{z_1\bar{z}_1}\left[k \right]\theta-2
\end{array}
\right)-\left(
\begin{array}{cc}
1 & 0 \\
0 & 1
\end{array}
\right),
\end{align}
then $h\left[k \right]$ is an Einstein (Ricci-flat) metric.
\end{prop}

A concrete example of $k$-instanton curvature tensors is given in \cite{Ishikawa:2001ye}
and the curvature is written by using linear operators on a Fock space.
It is known from (\ref{Fockrep}) and Proposition \ref{dictionary}
how to translate the operators into functions. (See also Appendix \ref{fock} and \cite{Sako:2016gqb}.)
Then the $k$-instanton curvature tensor is expressed by concrete elementary functions as follows:
\begin{align*}
\lefteqn{\hat{F}^-_{z_1\bar{z}_1}\left[k \right]=\frac{\mathrm{i}}{\zeta}-\frac{\mathrm{i}}{\zeta}\sum _{n_2=0}^\infty
 \frac{z_2^{n_2}\mathrm{e}^{-\frac{z^1\bar{z}^1+z^2\bar{z}^2}{\zeta}}\bar{z}_2^{n_2}}{n_2!\zeta^{n_2}} \left(d_1\left(0,n_2;k \right) \right)^2} \\
 &-\frac{\mathrm{i}}{\zeta} \sum _{n_1=1}^\infty \sum _{n_2=0}^\infty
\frac{z_1^{n_1}z_2^{n_2}\mathrm{e}^{-\frac{z^1\bar{z}^1+z^2\bar{z}^2}{\zeta}}\bar{z}_1^{n_1}\bar{z}_2^{n_2}}{n_1!n_2!\zeta^{n_1+n_2}}
\left\{\left(d_1\left(n_1,n_2;k \right) \right)^2-\left(d_1\left(n_1-1,n_2;k \right) \right)^2 \right\},
\end{align*}

\begin{align*}
\lefteqn{\hat{F}^-_{z_1\bar{z}_2}\left[k \right]=-\frac{\mathrm{i}}{\zeta}\frac{z_1^{k-1}z_2\mathrm{e}^{-\frac{z^1\bar{z}^1+z^2\bar{z}^2}{\zeta}}}
{\sqrt{\left(k-1 \right)!}\left( \sqrt{\zeta}\right)^{k}}d_1\left(k-1,1;k \right)d_2\left(0,0;k \right)} \\
 &-\frac{\mathrm{i}}{\zeta}\sum _{n_1=1}^{k-1} \frac{z_1^{n_1+k-1}z_2\mathrm{e}^{-\frac{z^1\bar{z}^1+z^2\bar{z}^2}{\zeta}}\bar{z}_1^{n_1}}
{\sqrt{\left(n_1+k-1 \right)!n_1!}\left( \sqrt{\zeta}\right)^{2n_1+k}}
 \left\{d_1\left(n_1+k-1,1;k \right)d_2\left(n_1,0;k \right)-d_1\left(n_1-1,0;k \right)d_2\left(n_1-1,0;k \right) \right\} \\
 &-\frac{\mathrm{i}}{\zeta} \sum _{n_1=1}^\infty \sum _{n_2=1}^\infty
\frac{z_1^{n_1-1}z_2^{n_2+1}\mathrm{e}^{-\frac{z^1\bar{z}^1+z^2\bar{z}^2}{\zeta}}\bar{z}_1^{n_1}\bar{z}_2^{n_2}}
{\sqrt{\left(n_1-1 \right)!\left(n_2+1 \right)!n_1!n_2!}\left( \sqrt{\zeta}\right)^{2n_1+2n_2}} \\
 &\times \left\{d_1\left(n_1-1,n_2+1;k \right)d_2\left(n_1,n_2;k \right)-d_1\left(n_1-1,n_2;k \right)d_2\left(n_1-1,n_2;k \right) \right\},
\end{align*}

\[\hat{F}^-_{z_1\bar{z}_2}\left[k \right] =- \hat{F}^-_{z_2\bar{z}_1}\left[k \right]^\dagger,\]
where $n_2\neq 0$ and
\begin{align}
\lefteqn{d_1\left(n_1,0;k \right)=\sqrt{n_1+k+1}\sqrt{\frac{\Lambda\left(n_1+k+1,0 \right)}{\Lambda\left(n_1+k,0 \right)}},}\nonumber \\
 &d_1\left(n_1,n_2;k \right)=\sqrt{n_1+1}\sqrt{\frac{\Lambda\left(n_1+1,n_2 \right)}{\Lambda\left(n_1,n_2 \right)}},\label{d1} \\
 &d_2\left(n_1,0;k \right)=\sqrt{\frac{\Lambda\left(n_1+k,1 \right)}{\Lambda\left(n_1+k,0 \right)}},\nonumber  \\
 &d_2\left(n_1,n_2;k \right)=\sqrt{n_2+1}\sqrt{\frac{\Lambda\left(n_1,n_2+1 \right)}{\Lambda\left(n_1,n_2 \right)}} \label{d2} .
\end{align}
Here
$$\Lambda\left[k \right]\left(n_1,n_2 \right)=\frac{w_k\left[k \right]\left(n_1,n_2 \right)}
{w_k\left[k \right]\left(n_1,n_2 \right)-2kw_{k-1}\left[k \right]\left(n_1,n_2 \right)},$$
and
$$w_n\left[k \right]\left(n_1,n_2 \right)=\sum _{l=0}^n \left\{ \frac{n!}{l!}\frac{\left(n_1-n_2+k+l \right)!}{\left(n_1-n_2-k \right)!}
\frac{2^{\left(n-l \right)}}{\left(n-l \right)!}\frac{\left(n_2+\left(n-l \right) \right)!}{n_2!}\right\}.$$
Note that some notations are slightly changed from \cite{Ishikawa:2001ye} and imaginary unit factor causes here.
See also Appendix \ref{imaginaryunit}.\\

Using these instanton curvatures, Hermitian-Einstein metrics
can be constructed by concrete elementary functions according to the Theorem \ref{masspro}.

\subsection{Einstein metric from finite $N$ }\label{finiten}
The full noncommutative $U\left(1 \right)$ instanton
solution is very complicated.
For simplicity, let us consider the $\zeta$-expansion. \\

In the previous subsection, $\hat{F}^-$ is represented by an infinite series
\begin{align}
\hat{F}^-=\sum _{n=1}^\infty\left(\frac{1}{\zeta} \right)^{\frac{n}{2}} \hat{F}^-_{\left({\frac{n}{2}} \right)}. \label{infiniteseries}
\end{align}
The anti-self-dual condition $\star \hat{F}^-=-\hat{F}^-$ implies
\begin{align}
\star \hat{F}^-_{\left({\frac{n}{2}} \right)}=-\hat{F}^-_{\left({\frac{n}{2}} \right)}
\end{align}
for each $n/2$.
Therefore it is possible to employ an arbitrary partial sum of (\ref{infiniteseries}) determined by a subset
$\displaystyle S\subset {\frac{1}{2}}\mathbb{Z}_{>0}$
\begin{align}
\hat{F}^-_S=\sum _{{\frac{n}{2}} \in S}\left(\frac{1}{\zeta} \right)^{\frac{n}{2}} \hat{F}^-_{\left({\frac{n}{2}} \right)}\label{Fseries}
\end{align}
for the anti-self-dual two-form to construct a Hermitian-Einstein metric $h$
without losing rigorousness.\footnote{One may choose even more loose condition
than (\ref{Fseries}). One can choose a different subset $S$ for each
$\hat{F}^-_{z_1\bar{z}_1},\hat{F}^-_{z_1\bar{z}_2}$ to obtain a Hermitian-Einstein metric.}
In the following we consider
\begin{align}
\hat{F}^-_{\left\{{\frac{N}{2}} \right\}}:=\sum _{n=1/2}^{{N}/{2}}
\left(\frac{1}{\zeta} \right)^{\frac{n}{2}} \hat{F}^-_{\left({\frac{n}{2}} \right)}.
\end{align}

\begin{ex}
First let us make the Ricci-flat metric $h\left[k \right]_{\left\{1 \right\}}$ from ${\hat{F}^-_\mathbb{C}\left[k \right]}_{\left\{1 \right\}}$.
The curvature tensor in this case is ${\hat{F}^-_\mathbb{C}\left[k \right]}_{\left\{1 \right\}}=\left(
\begin{array}{cc}
\frac{\mathrm{i}}{\zeta} & 0 \\
0 & -\frac{\mathrm{i}}{\zeta}
\end{array}
\right)$, and its determinant is
$\det\left[{\hat{F}^-_\mathbb{C}\left[k \right]}_{\left\{1 \right\}} \right]=\frac{1}{\zeta^2}.$

So the metric $h\left[k \right]_{\left\{1 \right\}}$ is given by
\begin{align}
h\left[k \right]_{\left\{1 \right\}} &:=\frac{1}{4~\det\left[{\hat{F}^-_\mathbb{C}\left[k \right]}_{\left\{1 \right\}} \right]\theta^2-1}\left(
\begin{array}{cc}
-4\mathrm{i}{\hat{F}^-_{z_1\bar{z}_1}\left[k \right]}_{\left\{1 \right\}}\theta-2
& -4\mathrm{i}{\hat{F}^-_{z_1\bar{z}_2}\left[k \right]}_{\left\{1 \right\}}\theta \\
-4\mathrm{i}{\hat{F}^-_{z_2\bar{z}_1}\left[k \right]}_{\left\{1 \right\}}\theta
& 4\mathrm{i}{\hat{F}^-_{z_1\bar{z}_1}\left[k \right]}_{\left\{1 \right\}}\theta-2
\end{array}
\right)-\left(
\begin{array}{cc}
1 & 0 \\
0 & 1
\end{array}
\right) \nonumber \\
 &=\frac{1}{1-4\zeta^{-2}\theta^2}\left(
\begin{array}{cc}
1-4\zeta^{-1}\theta+4\zeta^{-2}\theta^2 & 0 \\
0 & 1+4\zeta^{-1}\theta+4\zeta^{-2}\theta^2
\end{array}
\right)=\left(
\begin{array}{cc}
\frac{1-2\zeta^{-1}\theta}{1+2\zeta^{-1}\theta}  & 0 \\
0 & \frac{1+2\zeta^{-1}\theta}{1-2\zeta^{-1}\theta}
\end{array}
\right). \nonumber
\end{align}
This corresponds to the Euclidean metric essentially.
\end{ex}

\begin{ex}
Let us make a Ricci-flat metric $h\left[k \right]_{\left\{2 \right\}}$ from ${\hat{F}^-_\mathbb{C}\left[k \right]}_{\left\{2 \right\}}$.
From (\ref{f11k}),(\ref{f12k}),
\begin{align}
{\hat{F}^-_\mathbb{C}\left[k \right]}_{\left\{2 \right\}}& =\frac{\mathrm{i}}{\zeta} \left[
1-\frac{z_2\bar{z}_2}{\zeta} \left(d_1\left(0,1;k \right) \right)^2
 -\frac{z_1\bar{z}_1}{\zeta}\left\{\left(d_1\left(1,0;k \right) \right)^2-\left(d_1\left(0,0;k \right) \right)^2 \right\} \right] \left(
\begin{array}{cc}
1 & 0 \\
0 & -1
\end{array}
\right) \nonumber \\
 &-\frac{\mathrm{i}d_1\left(k-1,1;k \right)d_2\left(0,0;k \right)}{\zeta^{1+k/2}\sqrt{\left(k-1 \right)!}}\left(
\begin{array}{cc}
0 & z_1^{k-1}z_2 \nonumber\\
\bar{z}_1^{k-1}\bar{z}_2 & 0
\end{array}
\right).  \nonumber
\end{align}
 Then its determinant is
\begin{align}
\det\left[{\hat{F}^-_\mathbb{C}\left[k \right]}_{\left\{2 \right\}} \right] &
=\frac{1}{\zeta^2} \left[
1-\frac{z_2\bar{z}_2}{\zeta} \left(d_1\left(0,1;k \right) \right)^2
 -\frac{z_1\bar{z}_1}{\zeta}\left\{\left(d_1\left(1,0;k \right) \right)^2-\left(d_1\left(0,0;k \right) \right)^2 \right\} \right]^2 \nonumber \\
 &+ \frac{\left\{d_1\left(k-1,1;k \right)\right\}^2\left\{d_2\left(0,0;k \right) \right\}^2z_1^{k-1}z_2\bar{z}_1^{k-1}\bar{z}_2 }
 {\zeta^{2+k}\left(k-1 \right)!}.\nonumber
\end{align}
So the metric $h\left[k \right]_{\left\{2 \right\}}$ is given by
\begin{align}
h\left[k \right]_{\left\{2 \right\}} &:=\frac{1}{4\det\left[{\hat{F}^-_\mathbb{C}\left[k \right]}_{\left\{2 \right\}} \right]\theta^2-1}\left(
\begin{array}{cc}
-4\mathrm{i}{\hat{F}^-_{z_1\bar{z}_1}\left[k \right]}_{\left\{2 \right\}}\theta-2
& -4\mathrm{i}{\hat{F}^-_{z_1\bar{z}_2}\left[k \right]}_{\left\{2 \right\}}\theta \\
-4\mathrm{i}{\hat{F}^-_{z_2\bar{z}_1}\left[k \right]}_{\left\{2 \right\}}\theta
& 4\mathrm{i}{\hat{F}^-_{z_1\bar{z}_1}\left[k \right]}_{\left\{2 \right\}}\theta-2
\end{array}
\right)-\left(
\begin{array}{cc}
1 & 0 \\
0 & 1
\end{array}
\right), \nonumber
\end{align}
which can be calculated concretely though its expression becomes complex.
To simplify this we assume $k>3$, then
\begin{align}
h\left[k \right]_{\left\{2 \right\}}&
=\left\{ \frac{2}{1-4\det\left[{\hat{F}_\mathbb{C}^-\left[k \right]}_{\left\{2 \right\}} \right] \theta^2}-1\right\}\left(
\begin{array}{cc}
1 & 0 \\
0& 1
\end{array}
\right)
+\frac{4\mathrm{i}{\hat{F}^-_{z_1\bar{z}_1}\left[k \right]}_{\left\{2 \right\}}\theta}
{1-4\det\left[{\hat{F}_\mathbb{C}^-\left[k \right]}_{\left\{2 \right\}}  \right]\theta^2}\left(
\begin{array}{cc}
1 & 0 \\
0& -1
\end{array}
\right)  \nonumber \\
 &=\left\{ \frac{2}{1-4\theta^2\zeta^{-2}\left[
1-\frac{z_2\bar{z}_2}{\zeta} \left(d_1\left(0,1;k \right) \right)^2
 -\frac{z_1\bar{z}_1}{\zeta}\left\{\left(d_1\left(1,0;k \right) \right)^2-\left(d_1\left(0,0;k \right) \right)^2 \right\} \right]^2}-1\right\}\left(
\begin{array}{cc}
1 & 0 \\
0& 1
\end{array}
\right)  \nonumber \\
&-\frac{\frac{4\theta}{\zeta} \left[
1-\frac{z_2\bar{z}_2}{\zeta} \left(d_1\left(0,1;k \right) \right)^2
 -\frac{z_1\bar{z}_1}{\zeta}\left\{\left(d_1\left(1,0;k \right) \right)^2-\left(d_1\left(0,0;k \right) \right)^2 \right\} \right] }
{1-4\theta^2\zeta^{-2}\left[
1-\frac{z_2\bar{z}_2}{\zeta} \left(d_1\left(0,1;k \right) \right)^2
 -\frac{z_1\bar{z}_1}{\zeta}\left\{\left(d_1\left(1,0;k \right) \right)^2-\left(d_1\left(0,0;k \right) \right)^2 \right\} \right]^2}\left(
\begin{array}{cc}
1 & 0 \\
0& -1
\end{array}
\right).  \nonumber
\end{align}
\end{ex}

In next subsection, we discuss a Hermitian-Einstein metric obtained from
$1$-instanton solution.

%\begin{fact}If $h\left[k \right]:=\iota_{sym}\left\{
%2\left(E_4-\iota_{skew}\left(\hat{F}^-_{\mathbb{C}}\left[k \right] \right) \theta^+ \right)^{-1}-E_4 \right\}$
%then $h\left[k \right]$ is an Einstein metric.
%\end{fact}

%\begin{fact}
%\[h\left[k \right]=\frac{1}{4\left|\hat{F}_\mathbb{C}^- \right|\theta^2-1}\left(
%\begin{array}{cc}
%-4\mathrm{i}\hat{F}^-_{z_1\bar{z}_1}\theta-2 & -4\mathrm{i}\hat{F}^-_{z_1\bar{z}_2}\theta \\
%4\mathrm{i}\hat{F}^-_{z_2\bar{z}_1}\theta & 4\mathrm{i}\hat{F}^-_{z_1\bar{z}_1}\theta-2
%\end{array}
%\right)-\left(
%\begin{array}{cc}
%1 & 0 \\
%0 & 1
%\end{array}
%\right)\]
%\end{fact}

\subsection{Hermitian-Einstein metric from a 1-instanton}\label{1inst}

For the simplest example of the Hermitian-Einstein metric given in the previous discussion,
we describe a Hermitian-Einstein metric obtained from a single noncommutative $U(1)$ instanton.
Now we pay attention to low order terms.\\

For $k=1,\hat{F}^-_{\mathbb{C}}\left[1 \right]$ is
\begin{align*}
\lefteqn{\hat{F}^-_{z_1\bar{z}_1}\left[1 \right]=\frac{\mathrm{i}}{\zeta}-\frac{\mathrm{i}z_2\bar{z}_2}{\zeta^2} \left(d_1\left(0,1;1 \right) \right)^2
 -\frac{\mathrm{i}z_1\bar{z}_1}{\zeta^2}\left\{\left(d_1\left(1,0;1 \right) \right)^2-\left(d_1\left(0,0;1 \right) \right)^2 \right\}
 +\mathcal{O}\left(\zeta^{-3} \right) } \\
 &=\frac{\mathrm{i}}{\zeta}- \frac{2\mathrm{i}}{3}\frac{z_2\bar{z}_2}{\zeta^2}-\frac{\mathrm{i}z_1\bar{z}_1}{\zeta^2}
 \left\{\frac{5}{2}-\frac{4}{3} \right\}+\mathcal{O}\left(\zeta^{-3} \right)
 =\frac{\mathrm{i}}{\zeta}- \frac{\mathrm{i}}{6\zeta^2}\left(4z_2\bar{z}_2+7z_1\bar{z}_1\right)+\mathcal{O}\left(\zeta^{-3} \right) \\
 &\hat{F}^-_{z_1\bar{z}_2}\left[1 \right]=-\frac{\mathrm{i}z_2}{\zeta^{3/2}}\left(1-\frac{z_1\bar{z}_1}{\zeta} -\frac{z_2\bar{z}_2}{\zeta} \right)
d_1\left(0,1;1 \right)d_2\left(0,0;1 \right)+\mathcal{O}\left(\zeta^{-3} \right) \\
&=-\frac{2\mathrm{i}z_2}{3\zeta^{3/2}}\left(1-\frac{z_1\bar{z}_1}{\zeta} -\frac{z_2\bar{z}_2}{\zeta} \right)+\mathcal{O}\left(\zeta^{-3} \right) \\
&\hat{F}^-_{z_2\bar{z}_1}\left[1 \right]=-\frac{\mathrm{i}\bar{z}_2}{\zeta^{3/2}}\left(1-\frac{z_1\bar{z}_1}{\zeta} -\frac{z_2\bar{z}_2}{\zeta} \right)
d_1\left(0,1;1 \right)d_2\left(0,0;1 \right)+\mathcal{O}\left(\zeta^{-3} \right)  \\
&=-\frac{2\mathrm{i}\bar{z}_2}{3\zeta^{3/2}}\left(1-\frac{z_1\bar{z}_1}{\zeta} -\frac{z_2\bar{z}_2}{\zeta} \right)+\mathcal{O}\left(\zeta^{-3} \right)
\end{align*}
from (\ref{f11k}),(\ref{f121}). Then
\begin{align}
det\left[\hat{F}_\mathbb{C}^-\left[1 \right]_{\left\{2 \right\}}\right]
%&=\left(\hat{F}^-_{z_1\bar{z}_1}\left[1 \right]_{\left\{2 \right\}} \right)^2
%-\hat{F}^-_{z_1\bar{z}_2}\left[1 \right]_{\left\{2 \right\}}\hat{F}^-_{z_2\bar{z}_1}\left[1 \right]_{\left\{2 \right\}} \\
% &=-\frac{1}{\zeta^2}\left( 1- \frac{7z_1\bar{z}_1+4z_2\bar{z}_2}{6\zeta}\right)^2
% +\frac{4z_2\bar{z}_2}{9\zeta^3}\left(1-\frac{z_1\bar{z}_1}{\zeta} -\frac{z_2\bar{z}_2}{\zeta} \right)^2 \nonumber \\
 &=\frac{4z_2\bar{z}_2}{9\zeta^5}\left(\zeta-z_1\bar{z}_1-z_2\bar{z}_2\right)^2
-\frac{1}{36\zeta^4}\left( 6\zeta- 7z_1\bar{z}_1-4z_2\bar{z}_2\right)^2
\end{align}

From this $1$-instaoton curvature, the Hermitian-Einstein metric is given as
\begin{align*}
 &h \left[1 \right]_{\left\{2 \right\}}:=\frac{1}{4~\det\left[\hat{F}_\mathbb{C}^-\left[1 \right]\right]_{\left\{2 \right\}}\theta^2-1}\left(
\begin{array}{cc}
-4\mathrm{i}\hat{F}^-_{z_1\bar{z}_1}\left[1 \right]_{\left\{2 \right\}}\theta-2 & -4\mathrm{i}\hat{F}^-_{z_1\bar{z}_2}\left[1 \right]_{\left\{2 \right\}}\theta \\
-4\mathrm{i}\hat{F}^-_{z_2\bar{z}_1}\left[1 \right]_{\left\{2 \right\}}\theta & 4\mathrm{i}\hat{F}^-_{z_1\bar{z}_1}\left[1 \right]_{\left\{2 \right\}}\theta-2
\end{array}
\right)-\left(
\begin{array}{cc}
1 & 0 \\
0 & 1
\end{array}
\right) \\
 &=\frac{4}{1-4\left\{\frac{4z_2\bar{z}_2}{9\zeta^5}\left(\zeta-z_1\bar{z}_1-z_2\bar{z}_2\right)^2
-\frac{1}{36\zeta^4}\left( 6\zeta- 7z_1\bar{z}_1-4z_2\bar{z}_2\right)^2 \right\}\theta^2} \\
 &\times \left\{\frac{1}{2}\left(
\begin{array}{cc}
1 & 0 \\
0 & 1
\end{array}
\right)+
\frac{\theta}{\zeta}\left(
\begin{array}{cc}
1 & 0 \\
0 & -1
\end{array}
\right)+\frac{2\theta}{3\zeta^{3/2}}\left(
\begin{array}{cc}
0 & z_2 \\
\bar{z}_2 & 0
\end{array}
\right)+\frac{\theta}{6\zeta^2}\left(
\begin{array}{cc}
-4z_2\bar{z}_2-7z_1\bar{z}_1 & 0 \\
0 & 4z_2\bar{z}_2+7z_1\bar{z}_1
\end{array}
\right)
\right. \\ &\left.
+\frac{2\theta}{3\zeta^{5/2}}\left(
\begin{array}{cc}
0 & -z_2\left( z_1\bar{z}_1+z_2\bar{z}_2 \right) \\
-\bar{z}_2\left(z_1\bar{z}_1+z_2\bar{z}_2 \right) & 0
\end{array}
\right) \right\}-\left(
\begin{array}{cc}
1 & 0 \\
0 & 1
\end{array}
\right).
\end{align*}

%Note that each factor of $1/ \zeta^n$ also cause Ricci flat metric.

\section{K\"ahler structure and Bianchi identity}\label{sect5}

In this section we discuss the K\"ahler condition on the metric derived from (anti-)self-dual two-forms
of noncommutative $U(1)$ instantons. We will clarify this issue by illuminating
the duality between the K\"ahler geometry and $U(1)$ gauge theory claimed in \cite{inov}.

\subsection{K\"ahler geometry and $U(1)$ gauge theory}

Let $M$ be a two-dimensional complex manifold with a K\"ahler metric
\begin{equation}\label{c-metric}
    ds^2 = h_{i\bar{j}} (z, \overline{z})  dz^i d \overline{z}^{{j}},
\end{equation}
where local complex coordinates are given by $z^i = {x}^{2i} + \mathrm{i}~ x^{2i-1}, \; (i= 1, 2)$.
A K\"ahler manifold is described by a single function $K(z, \overline{z})$,
so-called K\"ahler potential, defined by
\begin{equation}\label{k-metric}
 h_{i\bar{j}} (z, \overline{z}) = \frac{\partial^2  K(z, \overline{z})}{\partial z^i \partial \overline{z}^{j}}.
\end{equation}
The K\"ahler potential is not unique but admits a K\"ahler transformation given by
\begin{equation}\label{kah-gauge}
K(z, \overline{z}) \to K(z, \overline{z}) + f(z) + \overline{f}(\overline{z})
\end{equation}
where $f(z)$ and $\overline{f}(\overline{z})$ are arbitrary holomorphic and anti-holomorphic functions.
Two K\"ahler potentials related by the K\"ahler gauge transformation \eq{kah-gauge} give rise
to the same K\"ahler metric \eq{k-metric}.

\begin{df}[K\"ahler form \cite{besse}]
Given a K\"ahler metric (\ref{c-metric}), the K\"ahler form is a fundamental closed two-form defined by
\begin{align}
  \Omega = \mathrm{i}~ h_{i\bar{j}} (z, \overline{z})  dz^i \wedge d \overline{z}^{j}. \label{ftwo-form}
\end{align}
\end{df}

Note that the K\"ahler form \eq{ftwo-form} can be written as
\begin{equation}\label{k-curvature}
    \Omega = d \mathcal{A} \qquad \mathrm{and} \qquad
    \mathcal{A} = \frac{\mathrm{i}}{2} (\overline{\partial} - \partial) K (z, \overline{z})
\end{equation}
where the exterior differential operator is given by $d = \partial + \overline{\partial}$ with
$\partial = dz^i \frac{\partial}{\partial z^i}$ and $ \overline{\partial}
= d \overline{z}^{i} \frac{\partial}{\partial \overline{z}^{i}}$.
Then the above K\"ahler transformation \eq{kah-gauge} corresponds to a gauge transformation
for the one-form $\mathcal{A}$ given by
\begin{equation}\label{k-gauge}
    \mathcal{A} \to \mathcal{A} + d \lambda
\end{equation}
where $\lambda = \frac{\mathrm{i}}{2} \big(\overline{f}(\overline{z}) - f(z) \big)$.
This implies that the one-form $\mathcal{A}$ corresponds to $U(1)$ gauge fields
or a connection of holomorphic line bundle.
Note that the K\"ahler form $\Omega$ on a K\"ahler manifold $M$ is a nondegenerate, closed two-form.
Therefore the K\"ahler form $\Omega$ is a symplectic two-form.
This fact leads to the following proposition:

\begin{prop}\label{kahler-symplectic}
A K\"ahler manifold $(M, \Omega)$ is a symplectic manifold too
although the reverse is not necessarily true.
\end{prop}

The K\"ahler condition enforces a specific analytic characterization of K\"ahler metrics:
\begin{lem}\label{normal-kgh}
$ds^2$ is K\"ahler if and only if it osculates to order 2 to the Euclidean metric everywhere.
\end{lem}
\noindent
The proof of this lemma can be found in \cite{griffiths-harris} (Griffiths-Harris, p.~107).
It means that the existence of normal holomorphic coordinates around each
point of $M$ is equivalent to that of K\"ahler metrics.

Let us consider an atlas $\{(U_\alpha, \varphi_\alpha)| \alpha \in I \}$ on
the K\"ahler manifold $M$ and denote the K\"ahler form $\Omega$ restricted
on a chart $(U_\alpha, \varphi_\alpha)$ as $\omega_\alpha \equiv \Omega|_{U_\alpha}$.
According to the Lemma \ref{normal-kgh}, it is possible to write the local K\"ahler form as
\begin{equation}\label{kahler-gauge}
\omega_\alpha = B + F_\alpha,
\end{equation}
where $B$ is the K\"ahler form of $\mathbb{C}^2$.
Since the two-form $F_\alpha$ must be closed due to the K\"ahler condition,
it can be represented by $F_\alpha = d A_\alpha$. % on $U_\alpha$.
Using Eq. \eq{k-curvature} and $F_\alpha = \omega_\alpha - B$,
the one-form $A_\alpha$ on $U_\alpha$ can be written as the form
\begin{equation}\label{local-one-form}
    A_\alpha = \frac{\mathrm{i}}{2} (\overline{\partial} - \partial) \phi_\alpha (z, \overline{z})
\end{equation}
where $\phi_\alpha (z, \overline{z}) = K_\alpha (z, \overline{z}) - K_0 (z, \overline{z})$ and
$K_\alpha (z, \overline{z})$ is the K\"ahler potential on a local chart $U_\alpha$ and $K_0 (z, \overline{z})
= z^i \overline{z}^{\bar{i}}$ is the K\"ahler potential of $\mathbb{C}^2$.
On an overlap $U_\alpha \bigcap U_\beta$, two one-forms $A_\alpha$ and $A_\beta$ can be glued
using the freedom \eq{k-gauge} such that
\begin{equation}\label{g-transf}
    A_\beta = A_\alpha + d \lambda_{\alpha\beta}
\end{equation}
where $\lambda_{\alpha\beta}(z, \overline{z})$ is a smooth function on the overlap $U_\alpha \bigcap U_\beta$.
The gluing \eq{g-transf} on $U_\alpha \bigcap U_\beta$ is equal to the K\"ahler transformation
\begin{equation}\label{holo-gauge}
K_\beta (z, \overline{z}) = K_\alpha (z, \overline{z})
+ f_{\alpha\beta} (z) + \overline{f}_{\alpha\beta} (\overline{z})
\end{equation}
if $\lambda_{\alpha\beta}(z, \overline{z}) = \frac{\mathrm{i}}{2} \big(\overline{f}_{\alpha\beta}(\overline{z})
- f_{\alpha\beta}(z) \big)$.

\begin{rem}\label{kahler-section}The K\"ahler transformation (\ref{holo-gauge}) implies the relation
\[e^{K_\beta} = | e^{f_{\alpha\beta}}|^2  e^{K_\alpha}.\]
So $e^{K(z, \overline{z})}$ is a section of a nontrivial line bundle over $M$.
\end{rem}

According to the proposition (\ref{kahler-symplectic}),
the K\"ahler manifold $(M, h)$ is also a symplectic manifold $(M, \Omega)$.
Therefore one can find a coordinate transformation $\varphi_\alpha \in \mathrm{Diff}(U_\alpha)$
on a local coordinate patch $U_\alpha$ such that $\varphi_\alpha^* (B + F) = B$
according to the famous Darboux theorem or Moser lemma in symplectic geometry \cite{symp-book}.
In other words, the electromagnetic fields in the local K\"ahler form \eq{kahler-gauge} can always be
eliminated by a local coordinate transformation.
To be specific, the Darboux theorem ensures the existence of the local coordinate transformation
$\varphi_\alpha: y^\mu \mapsto x^a = x^a (y), \; \mu, a = 1, \cdots, 4$, obeying \cite{hliu,sw-darboux}
\begin{equation}\label{darboux-tr}
    \Big (B_{ab} + F_{ab} (x) \Big) \frac{\partial x^a}{\partial y^\mu}
    \frac{\partial x^b}{\partial y^\nu}
    = B_{\mu\nu}.
\end{equation}
Note that $B_{ab}$ and $B_{\mu\nu}$ are constant since they are coming from the K\"ahler form
of $\mathbb{C}^2 \cong \mathbb{R}^4$ according to (\ref{kahler-gauge}) (see also the Lemma \ref{normal-kgh}).

\begin{rem}\label{gravity-darboux}So far the coordinates $x^\mu$ have been commonly used
for both gravity and field theory descriptions since it does not cause any confusion.
However it is convenient to distinguish two kinds of coordinates $(x^a, y^\mu)$
appearing in the Darboux transformation (\ref{darboux-tr}). The so-called Darboux coordinates $y^\mu$
will be used for field theory description while the so-called covariant coordinates $x^a$
will be used for gravity description.
\end{rem}

\begin{df}[Poisson bracket \cite{symp-book}]
Let $\theta := B^{-1} = \frac{1}{2} \theta^{\mu\nu}
\frac{\partial}{\partial {y}^\mu} \bigwedge \frac{\partial}{\partial y^\nu}
\in \Gamma(\Lambda^2 T\mathbb{R}^4)$ be a Poisson bivector.
Then the Poisson bracket $\{~\cdot~, ~\cdot~\}: C^\infty (\mathbb{R}^4) \times C^\infty (\mathbb{R}^4)
\to C^\infty (\mathbb{R}^4)$ is defined by $\{f,  g\} = \theta(df, dg)$
for any smooth functions $f, g \in C^\infty (\mathbb{R}^4)$.
\end{df}

Since both sides of Eq. \eq{darboux-tr} are invertible, one can take its inverse
and derive the following relation
\begin{equation}\label{darboux-map1}
    \Theta^{ab} (x) := \left(\frac{1}{B + F (x) } \right)^{ab}
    = \theta^{\mu\nu} \frac{\partial x^a}{\partial y^\mu}
    \frac{\partial x^b}{\partial y^\nu}
    = \{ x^a (y), x^b (y)\}
\end{equation}
or
\begin{equation}\label{darboux-map2}
    - \left(\frac{1}{1 + F\theta} B \right)_{ab} (x)
= \{\phi_a (y),  \phi_b (y)\}
\end{equation}
where $\phi_a (y) := B_{ab} x^b(y)$.
Recall that we have started with a K\"ahler manifold with the metric \eq{c-metric} and
applied the Darboux transformation to the local K\"ahler form \eq{kahler-gauge}.
Now, in the description \eq{darboux-map1} or \eq{darboux-map2}, the curving of the K\"ahler manifold
is described by local fluctuations of $U(1)$ gauge fields on the line bundle $L \to \mathbb{R}^4$.
This becomes more manifest by taking the coordinate transformation in Eq. \eq{darboux-tr} as the form
\begin{equation}\label{cov-phi}
    \phi_\mu (y) = p_\mu + a_\mu (y)
\end{equation}
and by calculating the Poisson bracket
\begin{equation}\label{poisson-br}
    \{\phi_\mu (y),  \phi_\nu (y) \}
    = -B_{\mu\nu} + \partial_\mu a_\nu(y) - \partial_\nu a_\mu (y)
    + \{a_\mu (y),  a_\nu (y) \} \equiv -B_{\mu\nu}
    + f_{\mu\nu} (y).
\end{equation}
The functions $a_\mu (y)$ in the Darboux transformation \eq{cov-phi} will be regarded
as gauge fields whose field strength is given by $f_{\mu\nu} (y) =  \partial_\mu a_\nu(y)
- \partial_\nu a_\mu (y) + \{a_\mu (y),  a_\nu (y) \}$.\footnote{Here $a_\mu$ is a gauge field of
a new $U(1)$ gauge symmetry with the Poisson structure rather than the original $U(1)$ gauge symmetry.
From the original $U(1)$ gauge theory point of view,
they are local sections of the line bundle $L \to \mathbb{R}^4$.}
Since they respect the non-Abelian structure due to the underlying Poisson structure,
they are different from ordinary $U(1)$ gauge fields $A_\mu (x)$ in \eq{local-one-form},
so they will be called ``symplectic" $U(1)$ gauge fields.
Then Eq. \eq{darboux-map2} leads to the exact Seiberg-Witten map between
commutative $U(1)$ gauge fields and symplectic $U(1)$ gauge fields \cite{sw-ncft,hliu,sw-darboux}:
\begin{equation}\label{esw-map}
    f_{\mu\nu} (y) =  \left(\frac{1}{1 + F \theta} F \right)_{\mu\nu} (x)
    \quad \mathrm{or} \quad F_{\mu\nu} (x) =  \left(\frac{1}{1 - f \theta} f \right)_{\mu\nu} (y).
\end{equation}
Thus the following Lemma is conferred \cite{hsy-jhep09,hsy-ijmpa09,hsy-ijmpa15}:

\begin{lem}\label{durboux-swmap}
The Darboux transformation $\varphi_\alpha \in \mathrm{Diff}(U_\alpha)$ on a local coordinate patch $U_\alpha$
obeying $\varphi_\alpha^* (B + F) = B$ is equivalent to the Seiberg-Witten map between
commutative $U(1)$ gauge fields and symplectic $U(1)$ gauge fields.
\end{lem}

The gauge theory description of K\"ahler gravity is realized by viewing a K\"ahler manifold
as a phase space and its K\"ahler form as the symplectic two-form on the phase space \cite{inov}.
This viewpoint naturally leads to a Poisson algebra $\mathfrak{P}=\{C^\infty(\mathbb{R}^4), \theta \}$
associated with the K\"ahler geometry we have started with.
The underlying Poisson structure is inherited from the symplectic structure, i.e.
$\theta = B^{-1} \in \Gamma(\Lambda^2 T\mathbb{R}^4)$,
which is a bivector field called the Poisson tensor. %i.e. a field of antisymmetric bilinear forms on $T^* \mathbb{R}^4$,

\subsection{K\"ahler metric and Bianchi identity}

Recall that the Seiberg-Witten map \eq{esw-map} has been derived from the
local K\"ahler form \eq{kahler-gauge}.
With the identification $\omega^\pm = f^\pm$ and using the map \eq{esw-map},
the metric $g^\pm$ in the Definition \ref{g-metric} can be written as
\begin{equation}\label{sw-metg}
g^\pm = 2 F^\pm \theta^\mp + E_4
\end{equation}
which can be inverted to yield
\begin{equation}\label{sw-comf}
F^\pm = \frac{1}{2} (g^\pm  - E_4) (\theta^\mp)^{-1}.
\end{equation}

Now we will prove the following proposition \cite{Lee:2012rb}.
\begin{prop}\label{kal-bianchi}%[S.~Lee, R.~Roychowdhury and H.~S.~Yang, \cite{Lee:2012rb}]
Let $F$ be a two-form in (\ref{sw-comf}).
Then the K\"ahler condition for the metric $g$ in (\ref{sw-metg}) is equivalent to
the Bianchi identity for the $U(1)$ curvature $f$.
\end{prop}

%\textcolor[rgb]{0.00,0.00,1.00}{Let us clarify why the Bianchi identity for commutative $U(1)$ gauge fields
%should be related to the one for noncommutative $U(1)$ gauge fields and
%it leads to the condition that the metric must be K\"ahler.}

\begin{pf}
First note that the K\"ahler condition for the metric $g$ in (\ref{sw-metg}) is the closedness
of the fundamental two-form $\omega = B + F$, which is equal to $d F = 0$.
Consider the Jacobi identity
\begin{equation}\label{x-jacobi}
\{ x^a,  \{ x^b, x^c \}  \}
+ \{ x^b,  \{ x^c, x^a \}  \}
+ \{ x^c,  \{ x^a, x^b \}  \} = 0
\end{equation}
that is equivalent to the Bianchi identity of symplectic $U(1)$ gauge fields
\begin{equation}\label{g-bianchi}
D_a f_{bc} + D_b f_{ca} + D_c f_{ab} = 0,
\end{equation}
where $D_a f_{bc} =\partial_a f_{bc}+ \left\{a_a,f_{bc} \right\}$.
Using Eq. \eq{darboux-map1}, let us rewrite the Jacobi identity \eq{x-jacobi} as
\begin{eqnarray}\label{x-bianchi}
0 &=& \{ x^a,  \Theta^{bc} (x)  \}_\theta
+ \{ x^b, \Theta^{ca} (x)  \}_\theta
+ \{ x^c,  \Theta^{ab} (x) \}_\theta \nonumber\\
&=& \theta^{\mu\nu}
\left( \frac{\partial x^a}{\partial y^\mu} \frac{\partial \Theta^{bc} (x)}
{\partial y^\nu}
+ \frac{\partial x^b}{\partial y^\mu}
\frac{\partial \Theta^{ca} (x)}{\partial y^\nu}
+ \frac{\partial x^c}{\partial y^\mu}
\frac{\partial \Theta^{ab} (x)}{\partial y^\nu} \right) \nonumber\\
&=& \theta^{\mu\nu}
\left( \frac{\partial x^a}{\partial y^\mu} \frac{\partial x^d}{\partial y^\nu}
\frac{\partial \Theta^{bc} (x)}{\partial x^d}
+ \frac{\partial x^b}{\partial y^\mu} \frac{\partial x^d}{\partial y^\nu}
\frac{\partial \Theta^{ca} (x)}{\partial x^d}
+ \frac{\partial x^c}{\partial y^\mu} \frac{\partial x^d}{\partial y^\nu}
\frac{\partial \Theta^{ab} (x)}{\partial x^d} \right) \nonumber\\
&=& \{ x^a, x^d \}_\theta \frac{\partial \Theta^{bc} (x)}{\partial x^d}
+ \{ x^b, x^d \}_\theta \frac{\partial \Theta^{ca} (x)}{\partial x^d}
+ \{ x^c, x^d \}_\theta \frac{\partial \Theta^{ab} (x)}{\partial x^d} \nonumber\\
&=&  \Theta^{ad} (x) \frac{\partial \Theta^{bc} (x)}{\partial x^d}
+  \Theta^{bd} (x) \frac{\partial \Theta^{ca} (x)}{\partial x^d}
+  \Theta^{cd} (x) \frac{\partial \Theta^{ab} (x)}{\partial x^d} \nonumber \\
&=& - \Theta^{ad} \Theta^{be} \Theta^{fc} \left(   \frac{\partial F_{ef} (x)}
{\partial x^d}
+ \frac{\partial F_{fd} (x)}{\partial x^e}
+  \frac{\partial F_{de} (x)}{\partial x^f} \right).
\end{eqnarray}
Since $\Theta^{ab}$ is invertible, we get from \eq{x-bianchi} the Bianchi identity
for the $U(1)$ curvature $F$, i.e.,
\begin{equation}\label{u1-bianchi}
 \frac{\partial F_{bc} (x)}{\partial x^a}
+ \frac{\partial F_{ca} (x)}{\partial x^b}
+  \frac{\partial F_{ab} (x)}{\partial x^c} = 0
  \qquad \Longleftrightarrow \qquad dF=0.
\end{equation}
The same argument shows that the reverse is also true, i.e., if $dF=0$,
the Bianchi identity (\ref{g-bianchi}) is deduced. This completes the proof.
\qed\end{pf}

If one introduces a new bivector $\Theta = \frac{1}{2} \Theta^{ab} (x) \frac{\partial}{\partial x^a}
\bigwedge \frac{\partial}{\partial x^b} \in \Gamma(\Lambda^2 T\mathbb{R}^4)$ using the Poisson tensor
in \eq{darboux-map1},
Eq. \eq{x-bianchi} shows that the Schouten-Nijenhuis bracket of
the bivector $\Theta \in \Gamma(\Lambda^2 TN)$ identically vanishes, i.e.,
$[\Theta, \Theta]_{SN} = 0$ \cite{vaisman}.
This means that the bivector $\Theta$ defines a new Poisson structure on $\mathbb{R}^4 \cong \mathbb{C}^2$.
We thus see that the Bianchi identity for symplectic $U(1)$ gauge fields leads to the Bianchi identity of commutative
$U(1)$ gauge fields and vice versa. Since the Bianchi
identity \eq{u1-bianchi} can be understood as the K\"ahler condition
for the local K\"ahler form \eq{kahler-gauge}, the Hermitian-Einstein metrics defined
by $g = \omega \cdot J$ must be K\"ahler.

Let us quantize the Poisson algebra $\mathfrak{P}$ to get a noncommutative algebra and
a corresponding noncommutative $U(1)$ gauge theory. We apply the deformation quantization $\mathcal{Q}$
in Appendix A and define the quantization map for symplectic $U(1)$ gauge fields \cite{ly-jkps2018}:
\begin{eqnarray}\label{q-map}
&& \mathcal{Q} (\phi_\mu) := \widehat{\phi}_\mu (y) = p_\mu + \widehat{A}_\mu (y), \nonumber \\
&& \mathcal{Q} (\{\phi_\mu, \phi_\nu \}) := - i [\widehat{\phi}_\mu (y), \widehat{\phi}_\nu (y)]
= -i \big(-B_{\mu\nu} + \widehat{F}_{\mu\nu}  (y) \big),
\end{eqnarray}
where $\mathcal{Q} (f_{\mu\nu}) :=
\widehat{F}_{\mu\nu}  (y) = \partial_\mu \widehat{A}_\nu (y)
- \partial_\nu \widehat{A}_\mu (y) - i [\widehat{A}_\mu (y), \widehat{A}_\nu (y)]$ is the field strength of
noncommutative $U(1)$ gauge fields $\widehat{A}_\mu (y) := \mathcal{Q} (a_\mu)$.
After quantization, the symplectic $U(1)$ gauge fields map to noncommutative $U(1)$ gauge fields
which contain infinitely many derivative corrections controlled
by the noncommutative parameter $\theta^{\mu\nu}$. For example,
the Seiberg-Witten map (\ref{esw-map}) receives noncommutative corrections
and takes a non-local form whose exact form was conjectured in \cite{hliu}:
\begin{equation}\label{esw-liu}
  F_{\mu\nu} (k) = \int d^4 y L_* \left[ \sqrt{1- \theta \widehat{F}}
  \left( \frac{1}{1-\widehat{F}\theta} \widehat{F} \right)_{\mu\nu} (y) W(y, C) \right] e^{ik \cdot y},
\end{equation}
where $W(x, C)$ is a straight open Wilson line, the determinant and rational function
of $\widehat{F}$ should be understood as a power series expansion,
and $L_*$ denotes the integrations together with the path ordering procedure.
The conjectured form (\ref{esw-liu}) was immediately proved in \cite{Okawa:2001mv,nc-openW}.
In a commutative limit where the derivatives of the field strength can be ignored, the map (\ref{esw-liu})
is reduced to the second form in (\ref{esw-map}).

An immediate question arises about the status of Proposition \ref{kal-bianchi}
after (deformation) quantization. Let us state the result with the following proposition.

\begin{prop}\label{cnc-bianchi}
Let $F$ be a two-form in (\ref{esw-liu}).
Then the closedness condition for the commutative $U(1)$ curvature $F$, $dF=0$,
is equivalent to the Bianchi identity for the noncommutative $U(1)$ curvature $\widehat{F}$.
\end{prop}

This proposition was proved in \cite{Okawa:2001mv} by proving the conjecture by H. Liu.
Theorem \ref{masspro} implies that the Hermitian metric $h^\pm$ in (\ref{masspro}) constructed
by the identification $\omega^\pm = \widehat{F}^\pm$ still generates a Ricci-flat metric.
Therefore Proposition \ref{kal-bianchi} may be lifted to noncommutative spaces although
we do not have a rigorous proof yet.

\section{Discussion}\label{sect6}

We have shown that the K\"ahler geometry can be described by a $U(1)$ gauge theory on a symplectic manifold
leading to a natural Poisson algebra associated with the K\"ahler geometry we have started with.
Since the Poisson algebra $\mathfrak{P}$ defined by the Poisson bracket
$\{f,g\} = \theta(df, dg)$ is mathematically
the same as the one in Hamiltonian dynamics of particles, one can quantize the Poisson algebra
in the exactly same way as quantum mechanics. Hence we have applied the deformation quantization
to the Poisson algebra $\mathfrak{P} = (C^\infty (\mathbb{R}^4), \{-, -\})$.
The quantization of the underlying Poisson algebra leads to a noncommutative
$U(1)$ gauge theory which arguably describes a quantized K\"ahler geometry,
as claimed in \cite{inov} and illuminated in \cite{ly-jkps2018}.
Then we get a remarkable duality between K\"ahler gravity and noncommutative $U(1)$ gauge theory
depicted by the following flow chart \cite{ly-jkps2018}:
\begin{equation} \label{q-diag}
\begin{array}[c]{ccc}
\mathrm{K\ddot{a}hler~gravity}&\stackrel{\mathfrak{I}^{-1}_\epsilon}{\longrightarrow}&
\mathrm{Symplectic~{\it U(1)}~gauge~theory }\\
{\mathcal{Q}}\downarrow\scriptstyle&&\downarrow{\mathcal{Q}}\scriptstyle\\
\mathrm{Quantized~K\ddot{a}hler~gravity} &\stackrel{\mathfrak{I}_\theta}{\longleftarrow}&
\mathrm{Noncommutative~{\it U(1)}~gauge~theory }
\end{array}
\end{equation}
Here $\mathcal{Q}: C^\infty (\mathbb{R}^4) \to \mathcal{A}_\theta$ means the quantization
and $\mathfrak{I}$ means an isomorphism between two theories.
In some sense $\mathfrak{I}$ corresponds to the gauge-gravity duality. It turns out \cite{ly-jkps2018} that
it can be interpreted as the large $N$ duality too. Since symplectic $U(1)$ gauge theory is
a commutative limit of noncommutative $U(1)$ gauge theory, we understand the classical isomorphism in \eq{q-diag}
as $\mathfrak{I}_\epsilon = \mathfrak{I}_\theta|_{\varepsilon = |\theta| \to 0}$.
The duality in \eq{q-diag} implies that a quantized K\"ahler gravity is isomorphically described by
a noncommutative $U(1)$ gauge theory. Actually this relation was already observed in \cite{inov}
in the context of topological strings probing K\"ahler manifolds where
several nontrivial evidences have been analyzed to support the picture.
In particular, the authors in \cite{inov} argue that
noncommutative $U(1)$ gauge theory is the fundamental description of K\"ahler gravity at all scales including
the Planck scale and provides a quantum gravity description such as quantum gravitational foams.
The duality in \cite{inov} has been further clarified in \cite{neova-kap} by showing that it follows
from the S-duality of the type IIB superstring.

This duality, if any, suggests an important clue about how to quantize the K\"ahler gravity.
Surprisingly, the correct variables for quantization are not metric fields but dynamical coordinates
$x^a(y)$ and their quantization is defined in terms of $\alpha'$
rather than $\hbar$. So far, there is no well-established clue to quantize metric fields directly
in terms of $\hbar$ in spite of impressive developments in loop quantum gravity.
However, the picture in \eq{q-diag} suggests a completely new quantization scheme
where quantum gravity is defined by quantizing spacetime itself in terms of $\alpha'$,
leading to a dynamical noncommutative spacetime described
by a noncommutative $U(1)$ gauge theory \cite{ly-jkps2018}.

The duality relation in \eq{q-diag} may be more accessible with the corresponding relation for solutions
of the self-duality equation, i.e., $U(1)$ instantons. Indeed it was shown in \cite{gi-u1-prl,gi-u1-plb,gi-u1-epl} that the commutative limit of noncommutative $U(1)$ instantons
are equivalent to Calabi-Yau manifolds.

{\bf Acknowledgments} \\
A.S. was supported in part by JSPS
KAKENHI Grant Number 16K05138.
The work of H.S.Y. was supported by the National Research Foundation of Korea (NRF) grant funded
by the Korea government (MOE) (No. NRF-2015R1D1A1A01059710) and (No. NRF-2018R1D1A1B07050113).

\appendix
\section{Deformation quantizations with separation of variables}\label{dq}

We summarize deformation quantization for Poisson manifolds and K\"ahler manifolds in Appendix \ref{dq}.

\begin{df}[Deformation quantization of Poisson manifolds]
Let $M$ be a Poisson manifold and $C^\infty \left(M \right)\left[\left[\zeta \right] \right]$ be a set of formal power series:
$C^\infty \left(M \right)\left[\left[\zeta \right] \right] := \left\{  f \ \Big| \
f = \sum_k f_k \zeta^k, ~f_k \in C^\infty \left(M \right)
\right\} .
$
A star product $*:C^\infty \left(M \right)\left[\left[\zeta \right] \right]\times C^\infty \left(M \right)\left[\left[\zeta \right] \right]
\rightarrow C^\infty \left(M \right)\left[\left[\zeta \right] \right]$ is defined as
\[f * g = \sum_k C_k (f,g) \zeta^k\]
such that the product satisfies the following (i)$\sim$(iv) conditions.
(i) $\left({C^\infty \left(M \right)\left[\left[\zeta \right] \right]},+,* \right)$ is a (noncommutative) algebra.
(ii) $C_k\left(\cdot ,\cdot  \right)$ is a bidifferential operator.
(iii) $C_0$ and $C_1$ are defined as
\begin{eqnarray}
&& C_0 (f,g) = f g,\qquad C_1(f,g)-C_1(g,f) = \{ f, g \}, \label{weakdeformation}
\end{eqnarray}
where $\{ f, g \}$ is the Poisson bracket of $M$.
(iv)$ f * 1 = 1 * f = f$.

$\left({C^\infty \left(M \right)\left[\left[\zeta \right] \right]},+,* \right)$ is called a deformation quantization of the Poisson manifold $M$.
\end{df}

Karabegov introduced a method to obtain a deformation quantization of a K\"ahler manifold in \cite{Karabegov1996}\cite{Karabegov}.
His deformation quantization is called deformation quantizations with separation of variables.
\begin{df}[A star product with separation of variables]
Let $*$ be a star product on a K\"ahler manifold as a Poisson manifold.
The $*$ is called a star product with separation of variables on a K\"ahler manifold when
\begin{eqnarray}
a * f = a f
\end{eqnarray}
for an arbitrary holomorphic function $a$ and
\begin{eqnarray}
f * b = f b
\end{eqnarray}
for an arbitrary anti-holomorphic function $b$.
\end{df}
The star product on $\mathbb{C}^2$
constructed by Karabegov's deformation quantization is given as
\begin{align}\label{f*g}
f*g=\sum _{n=0}^\infty \frac{\zeta^n}{n!}\delta^{k_1l_1}\cdots \delta^{k_nl_n}\left(\partial _{\bar{k}_1}\cdots \partial _{\bar{k}_n}f \right)
\left(\partial _{l_1}\cdots \partial _{l_n}g \right).
\end{align}

In this article we made Ricci-flat metrics from (anti-)self-dual two-forms on a noncommutative manifold.
A formal power series of symmetric two-form is not defined as a metric in ordinary sense.
For this reason we made Ricci-flat metrics from instantons on $\mathcal{F}$
instead of the noncommutative $\mathbb{C}^2$ described as a formal power series, in this article.
Here $\mathcal{F}$ is a noncommutative algebra constructed in Section \ref{sect4}
following the method in \cite{Sako:2016gqb}\cite{Sako:2017cix}.
The product of the algebra $\mathcal{F}$ is given by the star product (\ref{f*g}).
Then we obtain some Ricci flat metrics on $\mathbb{C}^2$.

\section{Noncommutative $U\left(1 \right)$ instanton in the Fock space} \label{imaginaryunit}
%The instantons are in \cite{Ishikawa:2001ye}.We will construct an Einstein metric from  noncommutative instantons.
In Appendix \ref{imaginaryunit}, we make a short review of the method to make a $U(1)$ instanton solution in \cite{Nekrasov:2000ih}
and multi instanton solutions in \cite{Ishikawa:2001ye}.\\

In noncommutative ${\mathbb R}^4$,
Nekrasov and Schwarz found
how to construct instanton gauge fields \cite{Nekrasov:1998ss}
by using the ADHM construction \cite{ADHM}.
Their work has encouraged studies of noncommutative ADHM instantons.
(See, for example, \cite{Ishikawa:2001ye,fu-kly,NCinstlecture}.)
Another method to construct noncommutative instantons
as smooth deformations of commutative instantons was provided in \cite{maeda_sako2,maeda_sako,sako4}.
The correspondence between the smooth deformation and the
ADHM construction are discussed in \cite{Maeda:2009qy}.
On the other hand, there exist instanton solutions
which are not smoothly connected
to commutative instantons.
The commutative limit of the noncommutative instantons are discussed in \cite{CommLimi,Hamanaka:2013tra}.

Noncommutative instantons are labeled by topological charge called instanton number.
The topological number of the noncommutative instanton is studied in
\cite{fu-kly,sako2,sako3,Tian:2002xw,furu-ptps}.
It is shown that the topological number coincides
with the dimension of a vector
space appearing in the ADHM construction.
In \cite{sako3}, this identification is shown when
the noncommutative parameter is self-dual for a $U(N)$ gauge theory.
In \cite{Tian:2002xw}, the equivalence is shown with
no restrictions on the noncommutative parameters, but
a noncommutative version of the Osborn's
identity (Corrigan's identity) is assumed.
In \cite{Hamanaka:2013tra} final version of 
the proof was announced.\\

In Definition \ref{cov}, a covariant derivative and gauge curvature are given as follows.
Covariant derivatives for scalar field $\phi\in \mathcal{F}$ on noncommutative $\mathbb{C}^2$ are defined as
$\hat{\nabla }_{z_l}\hat{\phi}:=\left[\hat{\partial }_{z_l},\hat{\phi }\right]+\hat{A}_{z_l}\hat{\phi}
=-\hat{\phi}\hat{\partial }_{z_l}+\hat{D}_{z_l}\hat{\phi}$
where we define a local gauge field $\hat{A}_{z_l}\in \mathcal{F}$ and $\hat{D}_{z_l}:=\hat{\partial }_{z_l}+\hat{A}_{z_l}.$
The gauge curvature is defined as
$\hat{F}_{z_l\bar{z}_m}:=\mathrm{i}\left[\hat{\nabla }_{z_l},\hat{\nabla }_{\bar{z}_m} \right]
=-\frac{\delta_{lm}}{\zeta_l}+\mathrm{i}\left[\hat{D}_{z_l},\hat{D}_{\bar{z}_m} \right].$

Using this notation, we introduce the ADHM construction in the following.
\subsection{Noncommutative ADHM construction}
\begin{df}Let $B_1,B_2 \in \mathbb{C}^{k\times k},~I \in \mathbb{C}^{k\times N},~J \in \mathbb{C}^{N\times k}$ be matrices satisfying
\begin{align}\label{dadhm}
\mu_\mathbb{C}:=\left[B_1,B_2^\dagger \right]+IJ=0,\quad
\mu_\mathbb{R}:=\left[B_1,B_1^\dagger \right]+\left[B_2,B_2^\dagger \right]+II^\dagger-J^\dagger J=\left(\zeta_1+\zeta_2 \right) E_k.
\end{align}
These equations are called the deformed ADHM equations. 
Here $\zeta_1,~\zeta_2$ are noncommutative parameter in (\ref{ncp}).
\end{df}
Let $E_k \in \mathbb{C}^{k\times k}$ be a unit matrix. We put $\beta_1,\beta_2 \in \mathbb{C}^{k\times k},\tau\in \mathbb{C}^{k\times \left(2k+1 \right)},
\sigma\in \mathbb{C}^{\left( 2k+1\right)\times k},\mathfrak{D}\in \mathbb{C}^{\left( 2k+1\right)\times 2k}$ as
\begin{align*}
&\beta_j:=\frac{B_j}{\sqrt{\zeta_j}},\quad \tau:=\left(B_2-z_2E_k,B_1-z_1E_k,I \right),\quad \sigma:=\left(-B_1+z_1E_k,B_2-z_2E_k,I \right)^T \\
 &\mathfrak{D}^\dagger:=\left(
\begin{array}{c}
 \tau \\
 \sigma^\dagger
\end{array}
\right)=\left(
\begin{array}{ccc}
B_2-z_2 &B_1-z_1 & I \\
-B_1^\dagger+\bar{z}_1 &B_2^\dagger-\bar{z}_2 & J^\dagger
\end{array}
\right).
\end{align*}

The first step of the ADHM construction is solving the deformed ADHM equations (\ref{dadhm}).

The second step of the ADHM construction is solving the equation $\mathfrak{D}^\dagger *\Psi=0,~\Psi^\dagger *\Psi=1$.

The third step of the ADHM construction is constructing gauge field $\hat{A}$ as
$\hat{A}_{z_l}:=\Psi^\dagger *\partial _{z_l}\Psi~,\hat{A}_{\bar{z}_l}:=\Psi^\dagger *\partial _{\bar{z}_l}\Psi $
where $\Psi$ is a solution of the equations in the second step.

Then the curvature tensor $\tilde{F}_{z_l\bar{z}_m}$ constructed from $\hat{A}_{z_l},~\hat{A}_{\bar{z}_m}$ is self-dual that means
$\tilde{F}_{z_l\bar{z}_m}$ is an instanton curvature tensor.

For the $U(1)$ case, this construction process can be expressed more explicitly.

Assume
\begin{align}\label{delta}
&\Psi:=\left(
\begin{array}{c}
\psi_+  \\
\psi_-  \\
 \xi
\end{array}
\right)=\left(
\begin{array}{c}
\sqrt{\zeta_2}\left(\beta_2^\dagger-a_2 \right)v  \\
\sqrt{\zeta_1}\left(\beta_1^\dagger-a_1 \right)v  \\
\xi
\end{array}
\right) \\
&\hat{\Delta}:=\zeta_1\left(\beta_1-a_1^\dagger \right)\left(\beta_1^\dagger -a_1\right)
+\zeta_2\left(\beta_2-a_2^\dagger \right)\left(\beta_2^\dagger -a_2\right),
\end{align}
where $\xi\in \mathcal{F} ,~ v\in\mathbb{C}^k\otimes \mathcal{F} $. $\mathcal{F}$ is defined in (\ref{ketbra}), and
 $\left(\beta_l^\dagger-a_l \right)v:=\left(\beta_l^\dagger\otimes id-E_k\otimes a_l \right)v$, where $id$ is an identity mapping.

A vector space $\mathcal{H}$ is defined using (\ref{fsp}) as
\[\mathcal{H}:=span_{\mathbb{C}}\left(\Ket{0,0},\Ket{1,0},\Ket{0,1},\Ket{1,1},\Ket{2,2}\cdots  \right).\]

\begin{df}A linear operators $P$ on $\mathcal{H}$ is defined as
\[P:=I^\dagger\left( \exp\sum _\alpha\beta^\dagger_\alpha a^\dagger_\alpha\right)\Ket{0,0}G^{-1}
\Bra{0,0}\left( \exp\sum _\alpha\beta_\alpha a_\alpha\right)I, \]
where
\[G:=\Bra{0,0}\left( \exp\sum _\alpha\beta_\alpha a_\alpha\right)II^\dagger
\left( \exp\sum _\alpha\beta^\dagger_\alpha a^\dagger_\alpha\right)\Ket{0,0}.\]
\end{df}

\begin{fact}This linear operator is a projection operator, i.e., $PP=P$.\end{fact}

A proposition below is true.
\begin{prop}Let $\Psi,~\hat{\Delta}v,~\xi$ be ones given above in (\ref{delta}). Then,
\[\mathfrak{D}^\dagger\Psi=0,~\Psi^\dagger *\Psi=1
\Longleftrightarrow \hat{\Delta}v+I\xi=0,~v^\dagger\hat{\Delta}v+\xi^\dagger\xi =1. \]
\end{prop}

\begin{lem}\label{ops}The operator $S$ which satisfies $SS^\dagger =id,S^\dagger S=id-P$ exists.
Let $\Lambda$ be $id+I^\dagger \hat{\Delta}^{-1}I$. If we put
\begin{align}\label{lambda}
\xi=\Lambda^{-1/2}S^\dagger,~v=-\hat{\Delta}^{-1}I\xi
%\hat{\Delta}v+I\xi=0,v^\dagger\hat{\Delta}v+\xi^\dagger\xi =1
\end{align}
then
\begin{align}
\hat{\Delta}v+I\xi=0,~v^\dagger\hat{\Delta}v+\xi^\dagger\xi =1 .
\end{align}
\end{lem}
This lemma means, if we find $\Lambda^{-1/2}$ and $\hat{\Delta}^{-1}$, 
then we can find a solution.\\

We define operators $\hat{\partial }_{z_l}$ and $\hat{D}_{z_l}$ on $\mathcal{H}$ in Section \ref{sect4} as
\[\hat{\partial }_{z_l}:=\frac{\bar{z}_l}{\zeta_l},\qquad
\hat{D}_{z_l}:=\hat{\partial }_{z_l}+\hat{A}_{z_l}.\]

Noncommutative $U\left(1 \right)$ instanton curvature in the Fock space is also defined as
\[\tilde{F}_{z_l\bar{z}_m}:=\mathrm{i}\left[\hat{\partial }_{\bar{z}_m},\hat{A}_{z_l} \right]_*
-\mathrm{i}\left[\hat{\partial }_{z_l},\hat{A}_{z_{\bar{m}}} \right]_*
+\mathrm{i}\left[\hat{A}_l, \hat{A}_{\bar{m}}\right]_*.\]
Using $\hat{D}_{z_l},~\tilde{F}$ is rewritten as
\begin{align}\label{tildeF}
\tilde{F}_{z_l\bar{z}_m}=\mathrm{i}\left[\hat{D}_{z_l}, \hat{D}_{\bar{z}_m}\right]_*+\frac{\mathrm{i}\delta_{lm}}{\zeta_l}.
\end{align}

Assume $\hat{A}_{z_l}:=\Psi^\dagger *\partial _{z_l}\Psi ,~\hat{A}_{\bar{z}_l}:=\Psi^\dagger *\partial _{\bar{z}_l}\Psi $ then
\[\hat{D}_{z_l}=-\frac{1}{\zeta_l}\Psi^\dagger \bar{z}_l\Psi,\qquad \hat{D}_{\bar{z}_l}=-\frac{1}{\zeta_l}\Psi^\dagger z_l\Psi.\]

Direct calculations derive the following results.
\begin{thm}\label{dzl}If $\Lambda:=id+I^\dagger \hat{\Delta}^{-1}I,\xi=\Lambda^{-1/2}S^\dagger,v=-\hat{\Delta}^{-1}I\xi$ then
\[\hat{D}_{z_l}=-\frac{1}{\sqrt{\zeta_l}}S\Lambda^{-1/2}a_l\Lambda^{1/2}S^\dagger,\qquad 
\hat{D}_{\bar{z}_l}=\frac{1}{\sqrt{\zeta_l}}S\Lambda^{1/2}a_l^\dagger \Lambda^{-1/2}S^\dagger.\]
\end{thm}

\begin{thm}\label{fzz}If $\tilde{F}^-_{z_k\bar{z}_l}$ is given by 
(\ref{tildeF}) and $\hat{D}_{z_l},\hat{D}_{\bar{z}_l}$ are 
defined in Theorem \ref{dzl}, then
\begin{align*}
 &\tilde{F}^-_{z_1\bar{z}_1}\left[k \right]=\frac{\mathrm{i}}{\zeta_1}-\frac{\mathrm{i}}{\zeta_1}S\Lambda^{-\frac{1}{2}}a_1\Lambda^{\frac{1}{2}}S^\dagger
S\Lambda^{\frac{1}{2}}a_1^\dagger\Lambda^{-\frac{1}{2}}S^\dagger +\frac{\mathrm{i}}{\zeta_1}S\Lambda^{\frac{1}{2}}a_1^\dagger\Lambda^{-\frac{1}{2}}S^\dagger
S\Lambda^{-\frac{1}{2}}a_1\Lambda^{\frac{1}{2}}S^\dagger, \\
 &\tilde{F}^-_{z_2\bar{z}_2}\left[k \right]=\frac{\mathrm{i}}{\zeta_2}-\frac{\mathrm{i}}{\zeta_2}S\Lambda^{-\frac{1}{2}}a_2\Lambda^{\frac{1}{2}}S^\dagger
S\Lambda^{\frac{1}{2}}a_2^\dagger\Lambda^{-\frac{1}{2}}S^\dagger +\frac{\mathrm{i}}{\zeta_2}S\Lambda^{\frac{1}{2}}a_2^\dagger\Lambda^{-\frac{1}{2}}S^\dagger
S\Lambda^{-\frac{1}{2}}a_2\Lambda^{\frac{1}{2}}S^\dagger, \\
&\tilde{F}^-_{z_1\bar{z}_2}\left[k \right]=-\frac{\mathrm{i}}{\sqrt{\zeta_1\zeta_2}}S\Lambda^{-\frac{1}{2}}a_1\Lambda^{\frac{1}{2}}S^\dagger
S\Lambda^{\frac{1}{2}}a_2^\dagger\Lambda^{-\frac{1}{2}}S^\dagger +\frac{\mathrm{i}}{\sqrt{\zeta_1\zeta_2}}S\Lambda^{\frac{1}{2}}a_2^\dagger\Lambda^{-\frac{1}{2}}S^\dagger
S\Lambda^{-\frac{1}{2}}a_1\Lambda^{\frac{1}{2}}S^\dagger, \\
&\tilde{F}^-_{z_2\bar{z}_1}\left[k \right]=-\frac{\mathrm{i}}{\sqrt{\zeta_1\zeta_2}}S\Lambda^{-\frac{1}{2}}a_2\Lambda^{\frac{1}{2}}S^\dagger
S\Lambda^{\frac{1}{2}}a_1^\dagger\Lambda^{-\frac{1}{2}}S^\dagger +\frac{\mathrm{i}}{\sqrt{\zeta_1\zeta_2}}S\Lambda^{\frac{1}{2}}a_1^\dagger\Lambda^{-\frac{1}{2}}S^\dagger
S\Lambda^{-\frac{1}{2}}a_2\Lambda^{\frac{1}{2}}S^\dagger.
\end{align*}
\end{thm}
This curvature is an instanton curvature.

\subsection{$U(1)~k$-instanton in the noncommutative $\mathbb{C}^2$}\label{fock}
In this section we summarize $U(1)$ multi-instanton solutions on $\mathbb{C}^2$ in \cite{Ishikawa:2001ye}.
For simplicity, let us assume $\zeta_1=\zeta_2=:\zeta$.
\begin{df}Noncommutative instanton curvature in the noncommutative $\mathbb{C}^2$ is defined as
\[\hat{F}^-_{\mathbb{C}}\left[k \right]=\left(
\begin{array}{cc}
\hat{F}^-_{z_1\bar{z}_1}\left[k \right] & \hat{F}^-_{z_1\bar{z}_2}\left[k \right] \\
-\hat{F}^-_{z_2\bar{z}_1}\left[k \right] & -\hat{F}^-_{z_1\bar{z}_1}\left[k \right]
\end{array}
\right):=\left(
\begin{array}{cc}
\iota\left(\tilde{F}^-_{z_1\bar{z}_1}\left[k \right] \right) & \iota\left(\tilde{F}^-_{z_1\bar{z}_2}\left[k \right] \right) \\
\iota\left(-\tilde{F}^-_{z_2\bar{z}_1}\left[k \right] \right) & \iota\left(-\tilde{F}^-_{z_1\bar{z}_1}\left[k \right] \right)
\end{array}
\right)\]
where $\iota$ is defined in Definition \ref{iota}.
\end{df}
We choose
$$B_1=\sum _{l=1}^{k-1}\sqrt{2l\zeta}e_le_{l+1}^\dagger,~B_2=,0~,I=\sqrt{2k\zeta}e_k,~J=0$$
as a solution of the deformed ADHM equations (\ref{dadhm}).
Here
$$e_l^\dagger=\left(
\begin{array}{ccccc}
\delta_{1,l} &\delta_{2,l}  &\cdots  &\delta_{k-1,l} & \delta_{k,l}
\end{array}
\right).
$$
In this case, the operator $S^\dagger$ in Lemma \ref{ops} is given by
\begin{align}\label{sdagger}
S^\dagger=\sum _{n_1=0}^\infty \Ket{n_1+k,0}\Bra{n_1,0}+\sum _{n_1=0}^\infty \sum _{n_2=1}^\infty \Ket{n_1,n_2}\Bra{n_1,n_2}.
\end{align}

From Theorem \ref{fzz} and (\ref{sdagger}), a $U(1)~k$-instanton curvature in the noncommutative $\mathbb{C}^2$ is obtained as follows.
\begin{align}
\tilde{F}^-_{z_1\bar{z}_1}\left[k \right]=&\frac{\mathrm{i}}{\zeta}
-\frac{\mathrm{i}}{\zeta}\sum _{n_2=0}^\infty \Ket{0,n_2}\Bra{0,n_2}\left(d_1\left(0,n_2;k \right) \right)^2 \nonumber \\
 &-\frac{\mathrm{i}}{\zeta} \sum _{n_1=1}^\infty \sum _{n_2=0}^\infty
\Ket{n_1,n_2}\Bra{n_1,n_2}\left\{\left(d_1\left(n_1,n_2;k \right) \right)^2-\left(d_1\left(n_1-1,n_2;k \right) \right)^2 \right\},\\
\tilde{F}^-_{z_2\bar{z}_2}\left[k \right]=&-\tilde{F}^-_{z_1\bar{z}_1}\left[k \right] \\
\tilde{F}^-_{z_1\bar{z}_2}\left[k \right]=&-\frac{\mathrm{i}}{\zeta} \Ket{k-1,1}\Bra{0,0}d_1\left(k-1,1;k \right)d_2\left(0,0;k \right) \\
 &-\frac{\mathrm{i}}{\zeta}\sum _{n_1=1}^{k-1} \Ket{n_1+k-1,1}\Bra{n_1,0}
 \left\{d_1\left(n_1+k-1,1;k \right)d_2\left(n_1,0;k \right)-d_1\left(n_1-1,0;k \right)d_2\left(n_1-1,0;k \right) \right\}  \nonumber \\
 &-\frac{\mathrm{i}}{\zeta} \sum _{n_1=1}^\infty \sum _{n_2=1}^\infty
\Ket{n_1-1,n_2+1}\Bra{n_1,n_2}  \nonumber\\  &\qquad \times\left\{d_1\left(n_1-1,n_2+1;k \right)d_2\left(n_1,n_2;k \right)
-d_1\left(n_1-1,n_2;k \right)d_2\left(n_1-1,n_2;k \right) \right\}  \nonumber  \\
\tilde{F}^-_{z_2\bar{z}_1}\left[k \right]=&\tilde{F}^-_{z_1\bar{z}_2}\left[k \right]^\dagger ,
\end{align}
where $d_1\left(n_1,n_2;k \right)$ and $d_2\left(n_1,n_2;k \right)$ 
are given by (\ref{d1})-(\ref{d2}).

Next we change these curvature operators into functions on $\mathbb{C}^2$ 
using the isomorphism (\ref{Fockrep}).

\begin{align} \hat{F}^-_{z_1\bar{z}_1}\left[k \right]:=&\iota\left(\tilde{F}^-_{z_1\bar{z}_1}\left[k \right] \right) \\
 =&\frac{\mathrm{i}}{\zeta}-\frac{\mathrm{i}}{\zeta}\sum _{n_2=0}^\infty
 \frac{z_2^{n_2}\mathrm{e}^{-\frac{z_1\bar{z}_1+z_2\bar{z}_2}{\zeta}}\bar{z}_2^{n_2}}{n_2!\zeta^{n_2}} \left(d_1\left(0,n_2;k \right) \right)^2 \nonumber \\
 &-\frac{\mathrm{i}}{\zeta} \sum _{n_1=1}^\infty \sum _{n_2=0}^\infty
\frac{z_1^{n_1}z_2^{n_2}\mathrm{e}^{-\frac{z_1\bar{z}_1+z_2\bar{z}_2}{\zeta}}\bar{z}_1^{n_1}\bar{z}_2^{n_2}}{n_1!n_2!\zeta^{n_1+n_2}}
\left\{\left(d_1\left(n_1,n_2;k \right) \right)^2-\left(d_1\left(n_1-1,n_2;k \right) \right)^2 \right\}. \nonumber \\
\hat{F}^-_{z_2\bar{z}_2}\left[k \right]:=&\iota\left(\tilde{F}^-_{z_2\bar{z}_2}\left[k \right] \right)=-\hat{F}^-_{z_1\bar{z}_1}\left[k \right]
\end{align}

\begin{align} \hat{F}^-_{z_1\bar{z}_2}\left[k \right]:=&\iota\left(\tilde{F}^-_{z_1\bar{z}_2}\left[k \right] \right)  \nonumber \\
=&-\frac{\mathrm{i}}{\zeta}\frac{z_1^{k-1}z_2\mathrm{e}^{-\frac{z_1\bar{z}_1+z_2\bar{z}_2}{\zeta}}}
{\sqrt{\left(k-1 \right)!}\left( \sqrt{\zeta}\right)^{k}}d_1\left(k-1,1;k \right)d_2\left(0,0;k \right) \\
 &-\frac{\mathrm{i}}{\zeta}\sum _{n_1=1}^{k-1} \frac{z_1^{n_1+k-1}z_2\mathrm{e}^{-\frac{z_1\bar{z}_1+z_2\bar{z}_2}{\zeta}}\bar{z}_1^{n_1}}
{\sqrt{\left(n_1+k-1 \right)!n_1!}\left( \sqrt{\zeta}\right)^{2n_1+k}} \nonumber  \\
&\times \left\{d_1\left(n_1+k-1,1;k \right)d_2\left(n_1,0;k \right)-d_1\left(n_1-1,0;k \right)d_2\left(n_1-1,0;k \right) \right\}  \nonumber \\
 &-\frac{\mathrm{i}}{\zeta} \sum _{n_1=1}^\infty \sum _{n_2=1}^\infty
\frac{z_1^{n_1-1}z_2^{n_2+1}\mathrm{e}^{-\frac{z_1\bar{z}_1+z_2\bar{z}_2}{\zeta}}\bar{z}_1^{n_1}\bar{z}_2^{n_2}}
{\sqrt{\left(n_1-1 \right)!\left(n_2+1 \right)!n_1!n_2!}\left( \sqrt{\zeta}\right)^{2n_1+2n_2}}  \nonumber \\
 &\times \left\{d_1\left(n_1-1,n_2+1;k \right)d_2\left(n_1,n_2;k \right)-d_1\left(n_1-1,n_2;k \right)d_2\left(n_1-1,n_2;k \right) \right\},  \nonumber \\
\hat{F}^-_{z_2\bar{z}_1}\left[k \right]=&\iota\left(\tilde{F}^-_{z_2\bar{z}_1}\left[k \right] \right)=-\overline{\hat{F}^-_{z_1\bar{z}_2}\left[k \right]}
\end{align}
where $\overline{a}$ is a complex conjugate of $a$.
\bigskip

In order to obtain Ricci-flat metrics in Subsection \ref{finiten} and Subsection \ref{1inst},
we need the first three terms of the expansion 
for $\hat{F}^-_{\mathbb{C}}\left[k \right]$ in $\sqrt{\frac{1}{\zeta}}$ .
\begin{align}
\hat{F}^-_{z_1\bar{z}_1}\left[k \right]  =&\frac{\mathrm{i}}{\zeta}-\frac{\mathrm{i}z_2\bar{z}_2}{\zeta^2} \left(d_1\left(0,1;k \right) \right)^2
 -\frac{\mathrm{i}z_1\bar{z}_1}{\zeta^2}\left\{\left(d_1\left(1,0;k \right) \right)^2-\left(d_1\left(0,0;k \right) \right)^2 \right\} \nonumber  \\
 &+\mathrm{i}\frac{z_1\bar{z}_1z_2\bar{z}_2}{\zeta^3} \left(d_1\left(0,1;k \right) \right)^2
 +\mathrm{i}\frac{\left(z_2\bar{z}_2 \right)^2}{\zeta^3}\left(d_1\left(0,1;k \right) \right)^2
+\mathrm{i}\frac{\left( z_1\bar{z}_1\right)^2}{\zeta^3}\left\{\left(d_1\left(1,0;k \right) \right)^2-\left(d_1\left(0,0;k \right) \right)^2 \right\} \nonumber  \\
&+\mathrm{i}\frac{z_1\bar{z}_1z_2\bar{z}_2}{\zeta^3}\left\{\left(d_1\left(1,0;k \right) \right)^2-\left(d_1\left(0,0;k \right) \right)^2 \right\}
+\mathcal{O}\left(\zeta^{-4} \right), \label{f11k}\\
d_1\left(0,0;k \right)=&\sqrt{\frac{\left(k+1 \right)\Lambda\left(k+1,0 \right)}{\Lambda\left(k,0 \right)}},
d_1\left(1,0;k \right)=\sqrt{\frac{\left(k+2 \right)\Lambda\left(k+2,0 \right)}{\Lambda\left(k+1,0 \right)}},
d_1\left(0,1;k \right)=\sqrt{\frac{\Lambda\left(1,1 \right)}{\Lambda\left(0,1 \right)}}, \nonumber
\end{align}
and
\begin{align}\hat{F}^-_{z_1\bar{z}_2}\left[k \right]=&-\frac{\mathrm{i}}{\zeta\left( \sqrt{\zeta}\right)^{k}}\frac{z_1^{k-1}z_2}{\sqrt{\left(k-1 \right)!}}
\left(1-\frac{z_1\bar{z}_1}{\zeta} -\frac{z_2\bar{z}_2}{\zeta}+\mathcal{O}\left(\zeta^{-2} \right) \right)
d_1\left(k-1,1;k \right)d_2\left(0,0;k \right)  \nonumber \\
 &-\frac{\mathrm{i}}{\zeta\left( \sqrt{\zeta}\right)^{k}}\sum _{n_1=1}^{k-1} \frac{z_1^{n_1+k-1}z_2\bar{z}_1^{n_1}}
{\sqrt{\left(n_1+k-1 \right)!n_1!}\zeta^{n_1}}
\left(1-\frac{z_1\bar{z}_1}{\zeta} -\frac{z_2\bar{z}_2}{\zeta}+\mathcal{O}\left(\zeta^{-2} \right) \right) \\
&\times  \left\{d_1\left(n_1+k-1,1;k \right)d_2\left(n_1,0;k \right)-d_1\left(n_1-1,0;k \right)d_2\left(n_1-1,0;k \right) \right\}  \nonumber \\
 &-\frac{\mathrm{i}}{\zeta} \sum _{n_1=1}^\infty \sum _{n_2=1}^\infty
\frac{z_1^{n_1-1}z_2^{n_2+1}\bar{z}_1^{n_1}\bar{z}_2^{n_2}}{\sqrt{\left(n_1-1 \right)!\left(n_2+1 \right)!n_1!n_2!}\zeta^{n_1+n_2}}
\left(1-\frac{z_1\bar{z}_1}{\zeta} -\frac{z_2\bar{z}_2}{\zeta}+\mathcal{O}\left(\zeta^{-2} \right) \right)  \nonumber \\
 &\times \left\{d_1\left(n_1-1,n_2+1;k \right)d_2\left(n_1,n_2;k \right)-d_1\left(n_1-1,n_2;k \right)d_2\left(n_1-1,n_2;k \right) \right\}. \nonumber
\end{align}

It is useful to distinguish the cases for $k=1$ and $k>1$.
\begin{align}
k=1\Rightarrow \hat{F}^-_{z_1\bar{z}_2}\left[1 \right] =&-\frac{\mathrm{i}z_2}{\zeta^{3/2}}
\left(1-\frac{z_1\bar{z}_1}{\zeta} -\frac{z_2\bar{z}_2}{\zeta} \right)
d_1\left(0,1;1 \right)d_2\left(0,0;1 \right)+\mathcal{O}\left(\zeta^{-3} \right). \label{f121}\\
 k>1\Rightarrow \hat{F}^-_{z_1\bar{z}_2}\left[k \right]&=-\frac{\mathrm{i}z_1^{k-1}z_2}{\zeta\left( \sqrt{\zeta}\right)^{k}\sqrt{\left(k-1 \right)!}}
\left(1-\frac{z_1\bar{z}_1}{\zeta} -\frac{z_2\bar{z}_2}{\zeta} \right)d_1\left(k-1,1;k \right)d_2\left(0,0;k \right) \nonumber  \\
 &-\frac{\mathrm{i}z_1^{k}z_2\bar{z}_1}{\sqrt{k!}\zeta^2\left( \sqrt{\zeta}\right)^{k}}
\left\{d_1\left(k,1;k \right)d_2\left(1,0;k \right)-d_1\left(0,0;k \right)d_2\left(0,0;k \right) \right\} \nonumber  \\
 &-\frac{\mathrm{i}z_2^2\bar{z}_1\bar{z}_2}{\sqrt{2!}\zeta^3}
\left\{d_1\left(0,2;k \right)d_2\left(1,1;k \right)-d_1\left(0,1;k \right)d_2\left(0,1;k \right) \right\} +\mathcal{O}\left(\zeta^{-4} \right). \label{f12k}
\end{align}
%$\zeta$

Functions $\Lambda,d_1,d_2$ for $k=1$ are useful for Subsection \ref{1inst} :
\begin{align*}
\lefteqn{\Lambda\left[1 \right] \left(n_1,n_2 \right)=
 \frac{\omega_1\left(n_1,n_2 \right)}
 {\omega_1\left(n_1,n_2 \right)-2\omega_{0}\left(n_1,n_2 \right)}=
 \frac{2+n_1+n_2}{n_1+n_2}, } \\
 & d_1\left(n_1,0;1 \right)
 =\sqrt{n_1+2}\sqrt{\frac{\Lambda\left[1 \right]\left(n_1+2,0 \right)}{\Lambda\left[1 \right]\left(n_1+1,0 \right)}}
 =\sqrt{\frac{\left(4+n_1 \right)\left(1+n_1 \right)}{\left(3+n_1 \right)}}, \\
 & d_1\left(n_1,n_2;1 \right)
=\sqrt{n_1+1}\sqrt{\frac{\Lambda\left[1 \right]\left(n_1+1,n_2 \right)}{\Lambda\left[1 \right]\left(n_1,n_2 \right)}}
=\sqrt{ \frac{\left(n_1+1 \right)\left(3+n_1+n_2 \right)\left( n_1+n_2\right)} {\left(1+n_1+n_2 \right)\left(2+n_1+n_2 \right)}}, \\
&d_2\left(n_1,0;1 \right)=\left\{\frac{\Lambda\left[1 \right]\left(n_1+1,1 \right)}{\Lambda\left[1 \right]\left(n_1+1,0 \right)} \right\}^{\frac{1}{2}}
%=\sqrt{\frac{\frac{2+n_1+2}{n_1+2}}{\frac{2+n_1+1}{n_1+1}}}
=\sqrt{\frac{\left(n_1+4 \right)\left(n_1+1 \right)}{\left(n_1+2 \right)\left(n_1+3 \right)}},  \\
&d_2\left(n_1,n_2;1 \right)=\sqrt{\frac{\left(n_2+1 \right)\Lambda\left[1 \right]\left(n_1,n_2+1 \right)}{\Lambda\left[1 \right]\left(n_1,n_2 \right)}}
%=\sqrt{\frac{\left(n_2+1 \right) \frac{3+n_1+n_2}{n_1+n_2+1}}{ \frac{2+n_1+n_2}{n_1+n_2}}}
=\sqrt{\frac{\left(n_2+1 \right)\left(n_1+n_2 \right)\left(3+n_1+n_2 \right)}{\left(n_1+n_2+1 \right)\left(2+n_1+n_2 \right)}}.%  \\
% &\Lambda\left[2 \right]\left(n_1,n_2 \right)=\frac{\omega_2\left(n_1,n_2 \right)}{\omega_2\left(n_1,n_2 \right)-4\omega_1\left(n_1,n_2 \right)} \\
%&=\frac{4+3n_1+n_1^2+5n_2+n_1n_2+n_2^2}{14-7k+k^2+7n_1-2kn_1+n_1^2+9n_2-2kn_2+n_1n_2+n_2^2-4\left(3-k+n_1+n_2 \right)}
\end{align*}

%%%%%%%%%%%%%%%%%%%%%%%%%%%%%%%

\end{document}